\documentclass[a4paper,twoside,twocolumn,english,showpacs]{revtex4}

\usepackage{times}
\usepackage[T1]{fontenc}
\usepackage{graphicx}
\usepackage{amsmath}
\usepackage{amssymb}
\usepackage{color}

\definecolor{MyLightMagenta}{cmyk}{0.5,0.9,0,0.5}

\makeatletter

\usepackage{babel}
\makeatother

\begin{document} 
\small 
\title{Theory of correlations between ultra-cold bosons released from an optical lattice}
\author{ E. Toth$^{1}$, A. M.  Rey$^{2}$ and P. B. Blakie$^{1}$}
\affiliation{$^{1}$ Jack Dodd Centre for Quantum Technology, Department of Physics, University of Otago, Dunedin, New Zealand}
\affiliation{$^{2}$ Institute for Theoretical Atomic, Molecular and Optical Physics, Cambridge, MA, 02138.}

\date{\today}


\begin{abstract}
 In this paper we develop a theoretical description of the correlations between ultra-cold bosons after free expansion from confinement in an optical lattice. We consider the system evolution during expansion and give criteria for a far field regime. We develop expressions for first and second order two-point correlations based on a variety of commonly used approximations to the many-body state of the system including Bogoliubov, meanfield decoupling, and particle-hole perturbative solution about the perfect Mott-insulator state. Using these approaches we examine the effects of quantum depletion and pairing on the system correlations.
Comparison with the directly calculated correlation functions is used to justify a Gaussian form of our theory from which we develop a general three-dimensional  formalism for inhomogeneous lattice systems  suitable for numerical calculations of realistic experimental regimes.

\end{abstract}
\maketitle
\section{Introduction}
In the past decade noise correlation analysis, analogous to the photon correlations observed in the landmark experiments of Hanbury-Brown and Twiss \cite{HBT}, has been applied to ultra-cold atom experiments.
In atomic systems such measurements can be used to reveal information about interaction-induced (i.e. many-body) correlations between the atoms. In addition, the dramatic differences between Bose and Fermi statistics have been clearly demonstrated with neutral atoms (e.g. see \cite{Jeltes2006a}).

A wide range of atomic physics experiments examining various aspects of correlations have been conducted. The first experiments  by Yasuda \emph{et al.} \cite{Yasuda1996a} observed atom bunching using a neutral beam of  (bosonic) neon atoms. In quantum degenerate Bose gases local third-order correlations have  been inferred by measuring  three-body decay rates \cite{Burt1997a,Tolra2004a}, and  first order coherence has been studied using matter wave interference \cite{Hagley1999a,Hadzibabic2006a} and Bragg spectroscopy \cite{Stenger1999a,Richard2003a}. Of most interest for the investigation in this paper has been the recent  experimental progress in the spatially resolved measurement of second order correlations in both  bosonic and fermionic ultra-cold  gases \cite{Schellekens2005a,Ottl2005a,Jeltes2006a,Folling2005a,Greiner2005a,Rom2006a}. Two general approaches are used to make these measurements. One approach involves directly counting atoms \cite{Schellekens2005a,Ottl2005a,Jeltes2006a}, while the other uses absorption imaging to measure the density \cite{Folling2005a,Greiner2005a,Rom2006a,Spielman2007a,Guarrerar2008a}. The applications of these measurements have included Bose and Fermi gases in harmonic traps \cite{Schellekens2005a,Jeltes2006a,Ottl2005a,Greiner2005a}, and in optical lattices \cite{Folling2005a,Rom2006a,Spielman2007a,Guarrerar2008a}.

In this work we are concerned with the theoretical formalism for the spatial noise correlations of  an ultra-cold Bose gas after expansion from an optical lattice, relevant to the experiments reported in Refs.~\cite{Folling2005a,Spielman2007a,Guarrerar2008a}. Initial theoretical work on this subject was provided by Altman \emph{et al.} \cite{Altman2004a}, who predicted the noise correlations using a perfect Mott-insulator approximation (i.e. neglecting tunneling) and assuming a simplified form for the single particle expansion.
Subsequent experiments in the Mott-insulator regime \cite{Folling2005a,Spielman2007a} verified those predictions, in particular the periodic bunching peaks in the noise correlations.
 Several recent theoretical proposals have built on that framework and investigated the use of noise correlations in characterizing  many-body states produced in optical lattices \cite{Rey2006a,Rey2006b,Rey2006c,Ashhab2005,Scarola2006,Hou2008}. This line of research provides an interesting new avenue for investigating the effects of interactions which compliments the other techniques available such as direct density imaging \cite{Greiner2002a,Gerbier2005,Spielman2007a}, Bragg \cite{Schori2004a,Stoferle2004a,Roth2004a,Batrouni2005a,Oosten2005a,Rey2005a,Blakie2002a} and Raman spectroscopy \cite{Gretchen2006a,Blakie2006d,Hazzard2007a}.

The basic organisation of the paper is as follows. In the remainder of this section we introduce the system of interest and give an introduction to how the far-field correlations are determined. In Sec.~\ref{SEC:SP} we derive the properties of single particle expansion from the lattice and use this to derive a simplified far field form and its validity conditions. We discuss the correlation function formalism and its relationship to the Bose-Hubbard Hamiltonian in Sec.~\ref{SEC:Corr}.  The main results for expanded correlations functions of a 1D lattice system are developed in Sec.~\ref{SEC:RESULTs} using a variety of theoretical approaches. In Sec.~\ref{SEC:GRESULTs} we introduce a Gaussian approach which we justify by comparison to the earlier results. In Sec.~\ref{SEC:3DRESULTs} we extend this Gaussian approach to a general 3D theory for the inhomogeneous lattice system, and then conclude.

\subsection{Optical lattice}
Consider a system of bosonic atoms in an optical lattice, described by the Hamiltonian
\begin{eqnarray}
H&=&\int d^3x\,\hat{\psi}^\dagger(\mathbf{x})\left(\frac{p^2}{2m}+\sum_{j=1}^3V_0\sin^2(kx_j)+V_{\rm{ext}}(\mathbf{x}) \right. \nonumber\\ &&+\left.\frac{U_0}{2}\hat{\psi}^\dagger(\mathbf{x})\hat{\psi}(\mathbf{x})\right)\hat{\psi}(\mathbf{x}),\label{FullH}
\end{eqnarray}
where $\hat{\psi}(\mathbf{x})$ is a bosonic field operator, $V_{\rm{ext}}$ describes any external potential (typically harmonic) and $U_0=4\pi a_s\hbar^2/m$ characterises the binary interactions between the particles, with $a_s$ the s-wave scattering length. The lattice is taken to be simple cubic, with $a=\pi/k$ and $b=2k$ the lengths of the direct and reciprocal lattice vectors along each direction, where $k$ is the wavelength of light used to produce the lattice. This lattice is hence of separable form, of the type produced in experiments with three sets of orthogonal counter-propagating light beams, and the \emph{depth} along each direction (i.e. $V_0$) is assumed to be the same.  The theory we present here can be easily extended to more general lattice configurations, but we restrict our attention to  this case for notational simplicity.
We define the quantities $\omega_R=\hbar k^2/2m$ and $E_R=\hbar
\omega_R$ as the recoil frequency and energy respectively.

\subsection{Measurement of expanded correlations}\label{MeasureIntro}
Due to the external confinement ($V_{\rm{ext}}$) the equilibrium state of this system will be of finite size $Ma$, where $M$ is an integer and $a$ is the lattice site spacing introduced above. Typically in experiments $M\lesssim 100$. It is of interest to probe the \emph{in situ} correlations of this system,  however imaging limitations make this difficult and it is necessary to turn off the lattice and external potentials and let the system expand first. Expansion itself is a rather complex process. The interactions between particles and high momentum components arising from the tight lattice confinement, mean that spontaneous $s$-wave scattering into unoccupied modes will generate additional noise and correlations in the system. Here we neglect these collisional processes and assume that upon release the atoms freely evolve \footnote{As future work the influence of these scattering events could be included, e.g. using the linearized formalism presented in Ref.~\cite{Molmer2008a}.}. After a short time of expansion the system density is sufficiently low that this is a good approximation, however this could be arranged experimentally by use of a Feshbach resonance.

Within this ideal expansion approximation the procedure for determining the expanded noise correlations is as follows: The atoms are initially in the optical lattice and are described  by some (interacting) manybody state $|\Phi\rangle$. The confining potentials are suddenly removed at $t=0$ and the system freely evolves according
to the propagator  $U(t)=e^{-iK t/\hbar}$, with $K=\int d^3x\,\hat{\psi}^\dagger(\mathbf{x})[p^2/2m]\hat{\psi}(\mathbf{x})$.
The mean system density at time
$t$ is given by
\begin{equation}
\langle \hat{n}(\mathbf{x}) \rangle_t \equiv \langle\Phi| U^{\dagger}(t)\hat\psi^
{\dagger}(\mathbf{x})\hat\psi(\mathbf{x})U(t)|\Phi\rangle.\label{mean_density}
\end{equation}
Experimentally the density can be measured using absorption imaging, however in any given measurement shot noise will be present and (\ref{mean_density}) corresponds to the average of many such measurements.
It has been demonstrated \cite{Greiner2002a,Spielman2007a} that under experimental conditions the atomic noise can be made to dominate the photon noise in the measurement process so that density correlations can be inferred from the noise analysis. In particular, the covariance $C(\mathbf{x},\mathbf{x}')=\langle\hat{n}(\mathbf{x}) \hat{n}(\mathbf{x}')\rangle-\langle\hat{n}(\mathbf{x})\rangle\langle \hat{n}(\mathbf{x}')\rangle$ can be measured using absorption images \footnote{In experiments  this observable  is integrated over the system volume to improve the signal strength.}, while the correlation function $G^{(2)}(\mathbf{x},\mathbf{x}')=\langle \hat{\psi}^{\dagger}(\mathbf{x}) \hat{\psi}^{\dagger}(\mathbf{x}^\prime)\hat{\psi}(\mathbf{x})\hat{\psi}(\mathbf{x}^\prime)\rangle$ is typically measured when direct atom counting is employed.

It is of interest to know the relationship of the noise correlations signal to the \emph{in situ} properties of the system. It is usually assumed that for large $t$ we can make
a \emph{far field} approximation whereby the density expectation
value becomes propotional to the momentum distribution of the initial
state, i.e.
\begin{equation}
\langle n(\mathbf{x}) \rangle \approx  \frac{m}{ht}\langle n_{\mathbf{Q}(\mathbf{x})} \rangle,\label{nFF}
\end{equation}
where
\begin{equation}\mathbf{Q}(\mathbf{x})=m\mathbf{x}/ht\label{Qdefn}
\end{equation}
relates the  \emph{in situ} momentum $\hbar \mathbf{Q}$ of
the field to the final observation position $\mathbf{x}$. Another important length scale we introduce at this point is the spatial separation of diffraction peaks which occur when the system is coherent. This length scale is set by the reciprocal lattice vector and is given by
\begin{equation}
l_b(t)=\hbar b t/m.\label{l_periodic}
\end{equation}
We also note the work of Kolovsky \cite{Kolovsky2004a}, who has performed numerical simulations of the second order correlation function for a small 1D lattice (in the pure Bose-Einstein condensate and perfect Mott-insulator limits) and observes interesting changes between the near-field and far-field correlations.

\section{Single particle expansion from a lattice}\label{SEC:SP}
\subsection{Semiclassical derivation}\label{SEC:Semiclassical}
\begin{figure}
\includegraphics[width=1.5in,keepaspectratio]{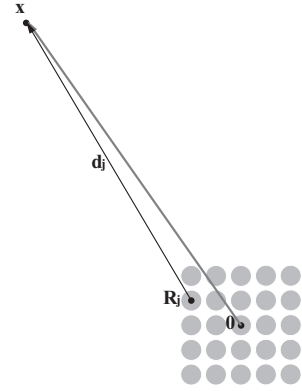}
\caption{\label{fig:expn_fig} Schematic of atomic far-field expansion  from a
lattice:
Consider an atom  initially $(t=0)$ located at lattice position $\mathbf{R}_{\mathbf{j}}$, 
and after expansion time $t$ is found at far-field position $\mathbf{x}$ (measured relative to the lattice center indicated by position $\mathbf{0}$). The initial lattice sites are indicated as grey filled-circles.}
\end{figure}
Consider Fig.~\ref{fig:expn_fig}: An atom initially in the lattice
at location $\mathbf{R}_{\mathbf{j}}$ expands in time $t$ to reach location
$\mathbf{x}$. We define the displacement vector for this particle during time-of-flight as $\mathbf{d}_{\mathbf{j}}\mathbf
{\equiv}\mathbf{x}-\mathbf{R}_{\mathbf{j}}$.
Observing that the average velocity of the particle must be $\mathbf{v}=\mathbf{d}_{\mathbf{j}}/t$, we can use the de Broglie relations to assign it a wavevector $\mathbf{Q}_{\mathbf{j}}=m\mathbf{d}_{\mathbf{j}}/\hbar t$, and respective
frequency  $\omega(\mathbf{Q}_
{\mathbf{j}})=\hbar |\mathbf{Q}_{\mathbf{j}}|^{2}/2m$.
Thus we conclude that the total phase change occurring to this particle
during expansion would be
\begin{equation}
\Delta\phi_{\mathbf{j}}=\mathbf{Q}_{\mathbf{j}}\cdot\mathbf{d}_{\mathbf{j}}-\omega(\mathbf{Q}_
{\mathbf{j}})t=\frac{m|\mathbf{d}_{\mathbf{j}}|^{2}}{2\hbar t}.\end{equation}
In the far-field approximation where $|\mathbf{x}|\gg |\mathbf{R}_{\mathbf{j}}|,$ we have $|\mathbf{d}_{\mathbf{j}}|^{2}
=|\mathbf{x}-\mathbf{R}_{\mathbf{j}}|^{2}\approx x^{2}-2\mathbf{x}\cdot\mathbf
{R}_{\mathbf{j}},$
so that \begin{equation}
\Delta\phi_{\mathbf{j}}\approx\Delta\phi-\mathbf{Q}\cdot\mathbf{R}_{\mathbf{j}},
\label{EQ:scphase}\end{equation}
where $\Delta\phi=\omega(\mathbf{Q})t$ and $\mathbf{Q}=m
\mathbf{x}/\hbar t$, as in Eq.~(\ref{Qdefn}).

We now turn to a quantum mechanical treatment of this expansion and show that we get the same spatially dependent phase factor (i.e. $-\mathbf{Q}\cdot\mathbf{R}_{\mathbf{j}}$)

\subsection{Wannier state expansion}\label{SEC:wannierexpn}
\begin{figure}
\includegraphics[width=2.2in,keepaspectratio]{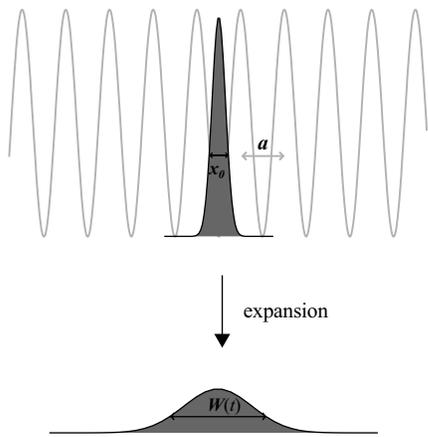}
\caption{\label{fig:expn2_fig} Far-field expansion of atoms from a
lattice:
A tight-binding Wannier state at lattice position $\mathbf{R}_{\mathbf{j}}$
expands for time $t$.}
\end{figure}
To construct an accurate quantum description it is necessary to consider the wavefunctions inside the lattice. Of most interest to us is the case where the atoms are in the ground band, and the  Wannier states, $w_0(\mathbf{x})$ form a convenient localized basis.  For sufficiently deep lattices a tight binding approximation is
appropriate, whereby we expanded the lattice potential to second order in the position variable about
each lattice site minimum to  obtain a local harmonic oscillator
description with frequency
\begin{equation}\omega_{\rm{Latt}}=2\sqrt{{V_0}/{E_R}}\,\omega_R.\label{wLatt}\end{equation}

For simplicity of argument, in this section we restrict our attention to the one-dimensional case of a lattice of $M$ sites, with locations
\begin{equation}
R_j=ja-\frac{1}{2}(M-1)a,\quad j=0,1,\ldots,M-1,
\end{equation}
and take $M$ to be odd for symmetry. Because the Wannier states are separable  our results and conclusions apply immediately along each direction in the 3D case.

The 1D Wannier states of the ground band can be approximated as
harmonic oscillator ground states
\begin{equation}
w_0 (x-R_j) \approx \frac{1}{\pi^{1/4}\sqrt{x_{0}}} e^{{- (x-R_j)
^2}/{2x_{0}^2}},
\end{equation}
where $x_{0}=\sqrt{\hbar/m\omega_{\rm{Latt}}}$ is the
oscillator length.
\begin{widetext}
After free evolution for time $t$ the harmonic oscillator approximation to the Wannier
state evolves to (see Fig.~\ref{fig:expn2_fig})
\begin{equation}
w_0(x-R_j,t)=\frac{1}{\pi^{1/4} \sqrt{W(t)}}\exp{\left\{\frac{-(x-R_j)
^2}{2[W(t)]^2}\right\}}\exp{\left\{i\frac{(x-R_j)^2}{2[W(t)]^2}\frac
{\hbar t}{mx_{0}^2}\right\}}e^{-i\theta},\label{FullWanExp}
\end{equation}
\end{widetext}
where $W(t)\equiv x_{0}\sqrt{1+\left( {\hbar t}/{mx_{0}^2}
\right)^2}$ is the time-dependent width of the Gaussian envelope,
and   $\theta\equiv \frac{1}{2}\arctan(t\hbar/mx_{0}^2)$.
Also see Refs.~\cite{Roth2003a,Bach2004a}.

\subsubsection{Far-field limit}
\begin{figure}
\includegraphics[width=3in,keepaspectratio]{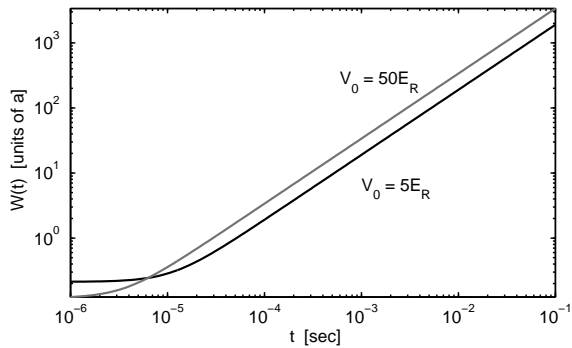}
\caption{\label{fig:wvpkt_fig}Evolution of Wannier state width $W(t)$
in units of the direct lattice vector $a$ for typical optical lattice
parameters and two lattice depths: $V_0=5E_R$ (black curve) and $V_0=50E_R$ (grey curve). Results for $^{87}$Rb in a lattice
made from counter-propagating $850$ nm light fields.}
\end{figure}
For sufficiently long expansion times the Wannier states will
expand  to be much larger than the total extent of the
 sites initially occupied in the optical lattice. We can then approximate
 \begin{equation}W(t)
\simeq\hbar t/mx_{0},\label{Wfarfield}\end{equation} i.e. the size of the system grows
linearly in time proportional to the initial momentum spread of the
Wannier state. We examine this expansion for typical experimental
parameters in Fig.~\ref{fig:wvpkt_fig}. We see that after about 10 ms
of expansion the individual Wannier wavepackets are of order 100
times the lattice site spacing. For the results we derive in this section, we assume that Eq.~(\ref{Wfarfield}) holds, but even more strictly that the system has expanded to be much larger than the spatial extent of the initially occupied lattice (i.e. $\max\{R_j\}\sim Ma$). Thus we have the hierarchy of length scales
\begin{equation}
W(t)\gg Ma\gg x_0.\label{fflengthscales}
\end{equation}
We also note that when Eq.~(\ref{Wfarfield}) is satisfied,   we have
\begin{equation}\frac{W(t)}{l_b(t)}=\frac{1}{2}\sqrt[4]{\frac{V_0}{E_R}} ,\end{equation}
so that the diffraction peaks will be further out than $W(t)$ unless $V_0>16E_R$.

The above length scales do not actually define the far-field limit. Instead, we define this limit as being when the following approximate form for the expanded Wannier states is accurate:
\begin{equation}
w_0(x-R_j,t)\simeq A(x,t) e^{-iQ(x)R_j},\label{EQ:w0}
\end{equation}
where the common (complex) amplitude of all Wannier states is
\begin{equation}
A(x,t) = \frac{\exp{\left\{\frac{-x^2}{2[W(t)]^2}\right\}}\exp\left\{i
\frac{\hbar Q(x)^2t}{2m}\right\}e^{-i\theta}}{\pi^{1/4}\sqrt{W(t)}},
\end{equation}
and $Q(x)$ is defined in Eq.~(\ref{Qdefn}).

We can consider this simplification as two distinct approximations to the phase and amplitude of the state, and we now consider the nature of these approximations. In particular, in addition to the requirements (\ref{fflengthscales}) we define the validity regime of the far-field regime in terms of the spatial region $X$  (i.e. far-field positions satisfying $|x|\ll X$) over which the amplitude approximation is accurate, and expansion time $T$ (i.e. $t\gg T$) for which the relevant phase errors will be small. The basic small parameter we will use is the ratio $R_j/W$ which has a maximum value of $Ma/W$.

\subsubsection{Amplitude approximation}
We first analyse the error in the amplitude of the Wannier state Eq.~(\ref{EQ:w0}) associated with ignoring the initial offset in the origin of the Wannier state (\ref{FullWanExp}),  i.e. in setting
\begin{equation}
|w_0(x-R_j,t)|\rightarrow
|A(x,t)|\equiv |w_0(x,t)|.
\end{equation}
To first order in $R_j/W$ the relative error in making this approximation is
\begin{equation}
\epsilon_A(R_j)=\frac{|R_jx|}{2W^2(t)}.\label{ErrA}
\end{equation}
To maintain a small amplitude error  we can restrict the spatial range over which Eq.~(\ref{EQ:w0}) is applied, i.e.
\begin{equation}
|x| \ll X_M(t)\equiv\frac{2W(t)^2}{Ma}.
\end{equation}
In practice  this means that the the amplitude approximation is accurate for $\max\{|x|\}\sim W$.

\subsubsection{Phase approximation}
We now consider the error in the phase of the Wannier state Eq.~(\ref{EQ:w0}) associated with ignoring the quadratic dependence on $R_j$ in the phase term of  the full Wannier state (\ref{FullWanExp}). To first order in $R_j/W$ the relative error in this approximation is
\begin{equation}
\epsilon_P(R_j) = \frac{R_j^2}{2W(t)x_0}.\label{ErrB}
\end{equation}
 Maintaining a small phase error between a pair of Wannier states separated by  $p$-sites requires that the expansion time satisfies
\begin{equation}
t\gg T_p\equiv\frac{p^2\pi^2}{4}\frac{1}{\omega_{R}}.\label{Eqtvalid}
\end{equation}
Satisfying this condition for all Wannier states (i.e. taking $p=M$) is difficult for the parameters of current experiments, and may require going beyond the $\exp(-iQR_j)$ approximation (\ref{EQ:w0}) to analyze the expanded images.

For the parameters used in Ref.~\cite{Folling2005a} where $M\sim80$ we have that $T_M\approx0.8\,$s, whereas in the experiments an expansion time of $22\,$ms was used (also see Table \ref{expttable}).
However, because they only examined the (incoherent) Mott-insulator regime, only short range correlations are important. 
For this situation the relevant quantity is $T_1\approx0.12ms$, which gives the timescale for the approximate expression (\ref{EQ:w0}) to accurately predict the phase relationship between atoms expanding from neighboring lattice sites.
 In situations where long-range phase coherence is important it will be necessary to use much longer expansion times or go beyond (\ref{EQ:w0}) in the analysis of the results.
\begin{table}[htdp]
\caption{Time scales and length scales for the experimental parameters  for $^{87}$Rb with  $V_0=30E_R$, $a= 425\,nm$, and $M\approx80$.}
\label{expttable}
\begin{center}
\begin{tabular}{|c|c|c|c|c|c|c|}
\hline
$x_0$ & $Ma$  & $W(t)$ & $X_M(t)$ & $t$ & $T_1$ & $T_M$ \cr
\hline
$58\,nm$ & $34\,\mu m$  & $277\,\mu m$ & $4.5\, mm$ & $22\,ms$ & $0.12\,ms$ & $0.80\,s$ \cr
\hline
\end{tabular}
\end{center}
\label{default}
\end{table}%

\subsection{Quasi-momentum basis}
It is also useful to consider an approximate form for the time-dependent expansion of quasi-momentum basis states. These states are defined as a discrete Fourier transform of the Wannier states, i.e. as
\begin{equation}
\phi_q(x,t)\equiv \frac{1}{\sqrt{M}}\sum_{j=0}^{M-1}e^{iqR_j}w_0(x-R_j,t).
\end{equation}
 Making the far-field approximation (\ref{EQ:w0}) we obtain
\begin{equation}
\phi_q(x,t)\approx A(x,t)F_M\left(Q(x)-q\right),\label{phi0}
\end{equation}
where we have defined
\begin{equation}
F({Q})\equiv\frac{\sin(\frac{1}{2}M{Q}a)}{\sqrt{M}\sin(\frac{1}{2}{Q}a)}.
\end{equation}
\begin{figure}
\includegraphics[width=3in,keepaspectratio]{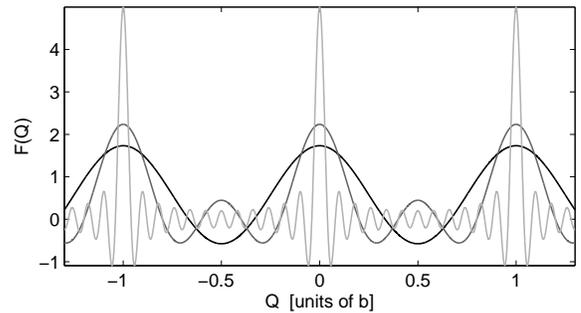}
\caption{\label{fig:FQM} The function $F(Q)$ for $M=3$ (black line), $M=5$ (medium grey line) and $M=25$ (light grey line).}
\end{figure}We note that $F(Q)$ is a peaked function and is periodic in $Q$ with period $b$. For $M$ large enough (see Fig.~\ref{fig:FQM}), $F(Q)$ is sharply peaked  at $Q=nb$ ($n=0,\pm1,\pm2,\ldots$) and with peak height of $\sqrt{M}$ and width $b/M$. Thus for large $M$ a useful approximation is to set $F(Q-q)\approx \sqrt{M}\delta_{Q,q+nb}$.  
From Eq.~(\ref{Qdefn}) we find that the spatial scale for this periodicity after expansion for time $t$ is $l_b$ as given in Eq.~(\ref{l_periodic}). The validity conditions for (\ref{phi0}) is the same as for the Wannier states discussed in the previous subsection.

For later convenience we will give several useful relations for these basis functions that can be easily derived,
\begin{eqnarray}
\sum_qF(Q-q)F(Q'\mp q)&=&\sqrt{M}F(Q\mp Q'),\label{Fsumrule}\\
F(Q-q)F(Q-q')&\approx&F(Q-q)^2\delta_{q,q'}.\label{Fdeltafn}
\end{eqnarray}
The second relation expresses the delta-function like property of the quasi-momentum basis functions, however for small lattices (i.e. small $M$) this approximation may not be appropriate.

\section{Correlation functions}\label{SEC:Corr}
\subsection{Quantum field operators and correlation functions}

The many-body state of the system is described by the Hamiltonian (\ref{FullH}) in terms of the bosonic quantum field
operator $\hat{\psi}(\mathbf{x})$ which obeys the usual equal time commutation
relations
\begin{equation}
\left[\hat{\psi}(\mathbf{x}),\hat{\psi}^{\dagger}(\mathbf{x}^\prime)\right] =
\delta (\mathbf{x}-\mathbf{x}^{\prime}), \qquad
\left[\hat{\psi}(\mathbf{x}),\,\hat{\psi}(\mathbf{x}^{\prime})\right] = 0.\label{EQ:psicomm}
\end{equation}
Correlation functions of the system can be expressed as expectations of products of the field operators at various positions and times (e.g. see \cite{Glauber1963a,Naraschewski1999a,Bezett2008a}).
Here we will mainly restrict ourselves to discussing normally ordered correlation functions
at a fixed time (i.e. after expansion).
The correlation functions we will be most interested in are the expanded density:
\begin{equation}
n(\mathbf{x})=\langle \hat{\psi}^{\dagger}(\mathbf{x})\hat{\psi}(\mathbf{x})\rangle,\label{expn}
\end{equation}
 the first order correlation correlation function (one-body density matrix):
\begin{equation}
G^{(1)}(\mathbf{x},\mathbf{x}^\prime)=\langle \hat{\psi}^{\dagger}(\mathbf{x})\hat{\psi}(\mathbf{x}^\prime)\rangle,\label{expG1}
\end{equation}
and the second order  correlation function:
\begin{equation}
G^{(2)}(\mathbf{x},\mathbf{x}^\prime)= \langle \hat{\psi}^{\dagger}(\mathbf{x})\hat{\psi}^{\dagger}(\mathbf{x}^\prime)
\hat{\psi}(\mathbf{x}^\prime)\hat{\psi}(\mathbf{x})\rangle,\label{expG2}
\end{equation}
(see Ref.~\cite{Naraschewski1999a}).
Normalized versions of these correlation functions can be defined as
\begin{eqnarray}
g^{(1)}(\mathbf{x},\mathbf{x}^\prime) &=& \frac{G^{(1)}(\mathbf{x},\mathbf{x}^\prime)}{\sqrt{ n(\mathbf{x})n(\mathbf{x}^\prime)}},\\
g^{(2)}(\mathbf{x},\mathbf{x}^\prime) &=& \frac{G^{(2)}(\mathbf{x},\mathbf{x}^\prime)}{{n(\mathbf{x})}{n(\mathbf{x}^\prime)}}.
\end{eqnarray}

Finally, we mention the density covariance function, defined as
\begin{equation}
C(\mathbf{x},\mathbf{x}^\prime)=\langle\hat{n}(\mathbf{x})\hat{n}(\mathbf{x}^\prime)\rangle-\langle\hat{n}(\mathbf{x})\rangle\langle\hat{n}(\mathbf{x}^\prime)\rangle,\label{Cov}
\end{equation}
where the density operator is $\hat{n}(\mathbf{x})=\hat{\psi}^{\dagger}(\mathbf{x})\hat{\psi}(\mathbf{x})$. This correlation function directly relates to the measurements made in experiments where density images are taken of the system and the shot noise analyzed. Commonly the normalized form of this observable is analyzed, defined as
\begin{equation}
\bar{C}(\mathbf{x},\mathbf{x}^\prime)=\frac{\langle\hat{n}(\mathbf{x})\hat{n}(\mathbf{x}^\prime)\rangle}{n(\mathbf{x})n(\mathbf{x}^\prime)}-1.
\end{equation}
To allow us to evaluate these correlation functions after expansion we now turn to developing an expression for the field operator in terms of the single particle modes discussed in the previous section.

\subsection{Bose-Hubbard Hamiltonian}
In the lattice the ultra-cold system  of bosons is well-described by the
tight-binding Bose-Hubbard model
\begin{equation}
\hat{H}_{\rm{BH}}=-J\sum_{\langle \mathbf{i}
,\mathbf{j}\rangle}\hat{a}^\dagger_{\mathbf{i}}\hat{a}_{\mathbf{j}}+\frac{1}{2}U\sum_{\mathbf{j}}\hat{a}_{\mathbf{j}}^\dagger\hat{a}^\dagger_{\mathbf{j}}\hat{a}_{\mathbf{j}}\hat{a}_{\mathbf{j}}+\sum_{\mathbf{j}}\epsilon_{\mathbf{j}}\hat{a}^\dagger_{\mathbf{j}}\hat{a}_{\mathbf{j}}, \label{BHH}
\end{equation}
where $J$ and $U$ are respectively the tunneling matrix element between sites
and the onsite interaction \cite{Jaksch1998a}, and  $\epsilon_{\mathbf{{j}}}$ is the local potential offset of each lattice
site (i.e. $\epsilon_{\mathbf{j}}=V_{\rm{ext}}(\mathbf{R}_{\mathbf{j}})$). The operator $\hat{a}^\dagger_{{\mathbf{j}}}$ creates a boson in a ground vibrational  state at
site ${\mathbf{j}}=\{j_x,j_y,j_z\}$ (i.e. at location $\mathbf{R}_{{\mathbf{j}}}$) of the lattice, and $\langle \mathbf{i},\mathbf{j}\rangle$ in the
summation indicates that the sum is restricted to nearest neighbours.

Eq.~(\ref{BHH}) can be derived from full Hamiltonian (\ref{FullH}) by considering the field operator projected to the ground band, i.e.
\begin{equation}
\hat{\psi}_g(\mathbf{x})=\sum_{\mathbf{j}}\hat{a}_{\mathbf{j}}w_0(\mathbf{x}-\mathbf{R}_{\mathbf{j}}),\label{psiBH}
\end{equation}
justified by the assumption that temperature and interaction effects are sufficiently small to remove the need to include modes of higher bands (e.g. see Refs.~\cite{Jaksch1998a,Blakie2004a}). In what follows we assume that prior to expansion the system is  in a  many-body eigenstate of  Eq.~(\ref{BHH}). Of particular interest is that this ground state undergoes a quantum phase transition as the values of $U$ and $J$ change, for instance, by changing the lattice depth. For the ratio $U/J$ below a critical value, $g_c$, the state of the system is superfluid with long range phase coherence. For $U/J>g_c$ the system enters the Mott insulating phase with a gapped excitation spectrum, suppressed number fluctuations and no long range phase coherence. In general $g_c$ depends on the dimensionality and the mean number of atoms per site.

\subsection{Field operators in the far-field limit}
 We now consider the many-body expansion of the system when the external potentials are released at $t=0$.
In the Heisenberg picture, the free evolution of the ground band field operator is given by
\begin{eqnarray}
\hat{\psi}_g(\mathbf{x},t) &=&\sum_{\mathbf{j}}\hat{a}_{\mathbf{j}}(0)w_0(\mathbf{x}-\mathbf{R}_{\mathbf{j}},t), \label{EQpsiGBt}
\end{eqnarray}
where we have taken the initial state to be entirely contained in the ground band (c.f Eq.~(\ref{psiBH})). Due to our approximation that the system freely evolves the time-dependence is completely contained in the single particle evolution (see Sec.~\ref{SEC:wannierexpn}).

Under the condition that initially only the ground band  is occupied, our expression for the field operator projected into the ground band (\ref{EQpsiGBt}) provides a useful and accurate description. However, care must be taken when evaluating correlation functions, as the commutation relation for $\hat{\psi}_g(\mathbf{x})$ is non-local, i.e.
\begin{equation}
[\hat{\psi}_g(\mathbf{x},t) ,\hat{\psi}_g^\dagger(\mathbf{x}^\prime,t)]=\sum_{\mathbf{j}}
w_0( \mathbf{x}-\mathbf{R}_{\mathbf{j}},t)w_0( \mathbf{x}^\prime-\mathbf{R}_{\mathbf{j}},t).
\end{equation}
For this reason it is more useful to work with normally ordered correlation functions, i.e. where annihilation operators precede creation operators. Indeed, $n(\mathbf{x})$, $G^{(1)}(\mathbf{x},\mathbf{x}^\prime)$ and $G^{(2)}(\mathbf{x},\mathbf{x}^\prime)$ take the same form (i.e. Eqs.~(\ref{expn})-(\ref{expG2})) whether we use the full or ground band projected fields (assuming only ground band modes are occupied in the manybody state).
On the other hand, the expression for the density covariance function in terms of $\hat{\psi}_g$ takes a different form compared to Eq.~(\ref{Cov}) due to the non-local commutation relation (see Appendix \ref{secprojcovprops}). However, it is more easily evaluated
as
\begin{equation}
C(\mathbf{x},\mathbf{x}^\prime)=G^{(2)}(\mathbf{x},\mathbf{x}^\prime)+n(\mathbf{x})\delta(\mathbf{x}-\mathbf{x}^\prime).\label{CovReln}
\end{equation}
 In the above expression the normally ordered $G^{(2)}(\mathbf{x},\mathbf{x}^\prime)$ and $n(\mathbf{x})$ functions can be evaluated using either $\hat{\psi}$ or $\hat{\psi}_g$. In the remainder of the paper we will concentrate on evaluating $n(\mathbf{x})$, $G^{(1)}(\mathbf{x},\mathbf{x}^\prime)$ and $G^{(2)}(\mathbf{x},\mathbf{x}^\prime)$  using the ground band projected operators. These results can be directly related to the density covariance using Eq.~(\ref{CovReln}) if needed.

Assuming that the expansion is sufficiently long for the far-field approximation to hold, we can express the expanded field operator in the simplified forms
\begin{eqnarray}
\hat{\psi}_g(\mathbf{x},t) &=& \sum_{\mathbf{j}} \hat{a}_{\mathbf{j}}A(\mathbf{x},t)e^{-i\mathbf{Q}(\mathbf{x})\cdot\mathbf{R}_{\mathbf{j}}},\label{psia}\\
&=& \sum_{\mathbf{q}} \hat{b}_\mathbf{q}A(\mathbf{x},t)F(\mathbf{Q}(\mathbf{x})-\mathbf{q}),\label{psib}
\end{eqnarray}
i.e.~in Wannier and Bloch mode expansions respectively, where $A(\mathbf{x},t)\equiv A(x,t)A(y,t)A(z,t)$ and $F(\mathbf{Q})\equiv F(Q_x)F(Q_y)F(Q_z)$ are the three-dimensional generalizations of the functions $A(x,t)$ and $F(Q)$ introduced in Sec.~\ref{SEC:SP}. 
 The operators $\hat{b}_{\mathbf{q}}$ are defined as the Fourier transform of the $\hat{a}_{\mathbf{j}}$ operators. We have suppressed time arguments on the mode operators which will be understood from hereon to take their $t=0$ (\emph{in situ}) value and remind the reader that $\mathbf{Q}(\mathbf{x})$ also has an implicit dependence on time as indicated in Eq.~(\ref{Qdefn}).

\section{Translationally invariant  results (1D)}\label{SEC:RESULTs}
In this section we give the two-point first and second order correlation functions for a system of bosons in a translationally invariant ($\epsilon_{\mathbf{j}}=0$) one-dimensional ($\mathbf{j}\to j$) optical lattice of $M=15$ sites for parameters corresponding to the superfluid and Mott-insulator phases. For consistency we will use the two sets of lattice parameters given in Table \ref{expttable2} to realizes these phases, where the expansion time, $t$, is chosen sufficiently long for the far-field approximation to hold. For ease of comparison we will always plot our correlation functions against $Q$ rather than $x$, since in terms of this variable of the correlation functions are  independent of expansion time in the far-field limit.
\begin{table}[htdp]
\caption{Time scales and length scales for the calculations presented in this paper. We take the case of $^{87}$Rb, with  $a= 425\,nm$, and $M=15$.}
\label{expttable2}
\begin{center}
\begin{tabular}{|l|c|c|c|c|c|c|c|}
\hline
& $V_0$  & $t$  & $T_M$ & $x_0$ & $Ma$  & $W(t)$ & $X_M(t)$  \cr
\hline
SF phase &  $9\,E_R$ & $50\, ms$  & $28\,ms$ & $78\, nm$ & $6.4\,\mu m$ & $465\,\mu m$ & $68\, mm$ \cr
\hline
MI phase & $18\,E_R$ & $50\, ms$  & $28\,ms$ & $66\,nm$ & $6.4\,\mu m$ & $554\,\mu m$ & $96\, mm$\cr
\hline
\end{tabular}
\end{center}
\label{default1}
\end{table}%

 For compactness of equations in what follows we abbreviate our notation according to:
 \begin{eqnarray}
 A(x,t)&\to &A, \\A(x',t)&\to& A',\\F(Q(x)-q)&\to& F_{Q-q}, \\ F(Q(x')-q)&\to& F_{Q'-q},
 \end{eqnarray} and so on.

\subsection{Pure Bose-Einstein condensate}\label{SECidSF}
\begin{figure}
\includegraphics[width=3.3in]{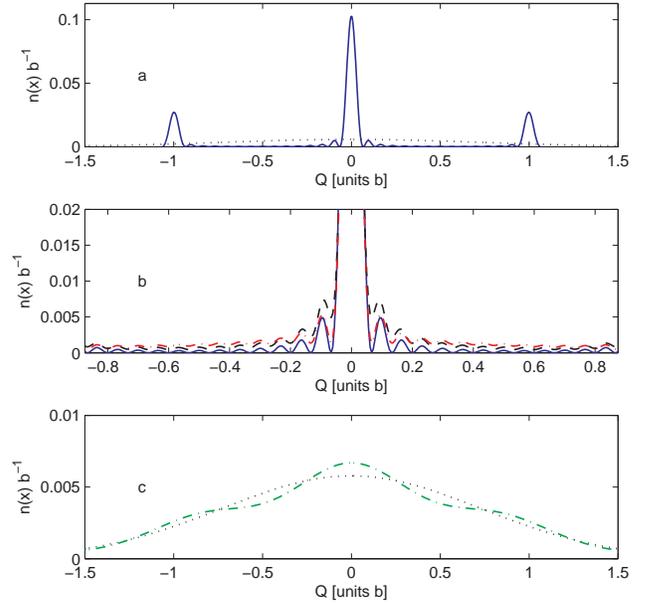}
\caption{\label{fig:IdealDen} 
Comparison of expanded density distributions.  (a) Perfect Mott-insulator (dotted line) and pure Bose-Einstein condensate (solid line) results. 
(b) Various superfluid phase results: pure Bose-Einstein condensate (solid line), Bogoliubov result (dashed line) and meanfield decoupling result (dash-dot line).
(c) Various Mott-insulator phase results:  perfect Mott-insulator result (dotted line) and including particle-hole corrections (dash-dot line).
Parameters:  $N\approx85$ atoms in a lattice with $M=15$. The Bogoliubov and meanfield decoupling calculations are for $U/2J\approx3.2$ and $N_0\approx65$ with non-condensate atom numbers being $\tilde{N}\approx20$ (Bogoliubov) and $\tilde{N}\approx17$ (decoupling). The perfect Mott-insulator and particle-hole results are for $U/2J\approx42$ with $N\approx85$ (see Table \ref{expttable2}).
}
\end{figure}

\begin{figure}
\includegraphics[width=3.3in]{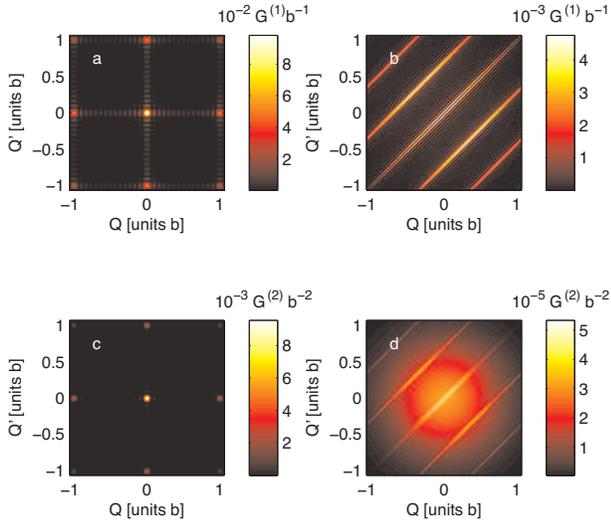}
\caption{\label{fig:IdealG1G2}
Two point first order correlation function, $G^{(1)}$, for  (a) pure Bose-Einstein condensate and (b) perfect Mott-insulator. Two point second order correlation function, $G^{(2)}$, for  (c) pure Bose-Einstein condensate and (d) perfect Mott-insulator. Parameters are the same as for the insulating cases considered in Fig.~\ref{fig:IdealDen}}
\end{figure}

The first approach we consider is the prototype superfluid: A \emph{pure Bose-Einstein condensate}. This approach is valid in the limit of vanishing interactions ($U=0$), but can also be used as an approximation for the interacting system in regimes where Gross-Pitaevskii theory is valid and quantum depletion can be neglected (see Sec.~\ref{SECbog}).
The ground state consists of all atoms in the $0$-quasimomentum mode, i.e.
\begin{equation}
|\Phi_{\rm{BEC}}\rangle =\frac{1}{\sqrt{N!}}\left(\hat{b}_0^\dagger\right)^N|0\rangle,
\end{equation}
where $N$ is the total number of atoms.
To calculate the correlation functions  we substitute the quasimomentum expansion (\ref{psib}) of the field operator into Eqs.~(\ref{expn})-(\ref{expG2}) for the correlations functions. Using that the only non-zero matrix elements are  $\langle \hat b^\dagger_0\hat b_0\rangle=N$ and  $\langle \hat
b^\dagger_0 \hat b^\dagger_0\hat b_0\hat b_0\rangle=N(N-1)$, we obtain
\begin{eqnarray}
n(x) &=& N|A|^2F_Q^2,\label{iSFn}\\
G^{(1)}(x,x') &=& NA^*A' F_QF_{Q'},\label{iSFG1}\\ \
G^{(2)}(x,x') &=& N(N-1)|A|^2|A'|^2 F_Q^2F_{Q'}^2.\label{iSFG2}
\end{eqnarray}
These expressions depend on the number of sites $M$ through the function $F_Q$. In Fig.~\ref{fig:IdealDen}(a), and Figs.~\ref{fig:IdealG1G2}(a) and (c) we show examples of these correlation functions for the pure Bose-Einstein condensate.

Result (\ref{iSFn}) is somewhat similar to the textbook example of  laser light passing through a diffraction grating. Long-range coherence is revealed through the constructive interference that gives rise to the density peaks (see Fig.~\ref{fig:IdealDen}(a)). These peaks occur where $Q$ is equal to a reciprocal lattice vector, as at these locations the phase accumulated by atoms propagating from distinct lattice sites are equal to within modulo $2\pi$. In terms of the coordinate $x$ the spacing between peaks is $l_b$ as given in Eq.~(\ref{l_periodic}).

The superfluid correlation functions are (approximately) separable in the sense that
\begin{eqnarray}
G^{(1)}(x,x')&=&\sqrt{n(x)n(x')},\\ G^{(2)}(x,x')&\approx& n(x)n(x'),
\end{eqnarray}
 (to order $1/N$), with the consequence that the normalized correlation functions are
 \begin{eqnarray}
 g^{(1)}(x,x')&=&\frac{A^*A'}{|A||A'|},\\  g^{(2)}(x,x')&=&1-1/N.
 \end{eqnarray}

\subsection{Perfect Mott-insulator}\label{SECpMI}
In the limit where   $U\gg J$, we can neglect the tunneling and approximate the ground state of the Bose Hubbard Hamiltonian (\ref{BHH}) as a \emph{perfect} Mott-insulator
\begin{equation}
|\Phi_{\rm{M}}\rangle = \prod^{M-1}_{j=0}\frac{1}{\sqrt{n!}} (\hat{a}^\dagger_j)^n|
0\rangle, \label{pmi}
\end{equation}
where  $n=N/M$ is the filling factor or mean number of particles per site. We have assumed that the number of particles is commensurate with lattice sites, i.e. that $n$ is an integer. Using the Wannier expansion (\ref{psib}) of the field operator to evaluate Eqs.~(\ref{expn})-(\ref{expG2}) we obtain
\begin{eqnarray}
 n_{\rm{M}} (x) &=&N|A|^2,\label{pMIn}\\
G_{\rm{M}} ^{(1)}(x,x') &=& \frac{N}{\sqrt{M}}A^*A'F_{Q-Q'},\label{pMIG1}\\
G_{\rm{M}} ^{(2)}(x,x') &=& N|A|^2|A'|^2\left[(N-n-1) +\frac{N}{M}F_{Q-Q'}^2\right],\label{pMIG2}
\end{eqnarray}
where we have used that 
\begin{eqnarray}
\langle\hat{a}_i^\dagger\hat{a}_j\rangle&=&n\delta_{ij},\\
\langle \hat{a}_i^\dagger\hat{a}_j^\dagger\hat{a}_k\hat{a}_l\rangle&=&n^2(\delta_{ik}\delta_{jl}+\delta_{il}\delta_{jk})-n(n+1)\delta_{ij}\delta_{kl}\delta_{jk},
\end{eqnarray}
and  $\sum_je^{i(Q-Q')R_j}=\sqrt{M}F_{Q-Q'}$. We have labelled these correlations with a subscript M for future reference. In Figs. 
\ref{fig:IdealDen}(a), and Figs.~\ref{fig:IdealG1G2}(b) and (d) we show examples of these correlation functions for the perfect Mott-insulator.

The expanded density of the Mott-insulator reveals the lack of phase coherence, i.e. the density arises as the incoherent sum of the individual Wannier states expanding from each site so that no interference peaks are observed (see Fig.~\ref{fig:IdealDen}(a)).

The correlation functions reveal many interesting features. Unlike the superfluid case, the Mott-insulator correlation functions are not separable. Indeed the correlation functions have well defined  ridges running down the diagonal ($Q=Q'$) and repeating periodically parallel to the diagonals (i.e. $Q=Q'+jb$ with $j=0\pm1,\pm2,\ldots$), see Figs.~\ref{fig:IdealG1G2}(b) and (d) .  The normalized first order correlation function is
\begin{equation}
g^{(1)}(x,x')=\frac{A^*A'}{|A||A'|}\frac{F_{Q-Q'}}{\sqrt{M}},
\end{equation} and reveals the absence of phase coherence with  off-diagonal decay over a distance scale of $\Delta x\sim ht/mM$.  

The second order correlations have a diagonal ridge similar to those in $G^{(1)}$, but also a smooth background envelope.
The normalized form is
\begin{equation}
g^{(2)}(x,x')\approx 1+|g^{(1)}(x,x')|^2,
\end{equation}
where we have neglected terms of order $n/N$ as being small. The second term (i.e. $|g^{(1)}(x,x')|^2$), reveals the usual tendency for bosonic particles to  cluster together.

 The results of this and the previous subsection emphasize the marked difference between the expanded correlations of the superfluid and Mott-insulator phases in an optical lattice.

\subsection{Finite temperature ideal gas}\label{SECftIG}
It is of interest to compare the Mott-insulator properties with those of an ideal gas at finite temperature \cite{Wild2006a,Blakie2004b,Blakie2007a,Blakie2007e}. For this system the quasimomentum modes form a diagonal basis, which fluctuate according to  
\begin{eqnarray}
\langle \hat{b}_{q1}^\dagger\hat{b}_{q2}\rangle&=&\bar{n}_{q1}\delta_{q1,q2},\\
 \langle \hat{b}_{q1}^\dagger \hat{b}_{q2}^\dagger \hat{b}_{q1} \hat{b}_{q2}\rangle&=&\bar{n}_{q1}\bar{n}_{q2}(1+\delta_{q1,q2}), 
 \end{eqnarray}
in the grand canonical description,  where $\bar{n}_q$ is the mean occupation given by the Bose-Einstein distribution.
 The correlation functions are given by
 \begin{eqnarray}
 n(x) &=&|A|^2\sum_q\bar{n}_qF_{Q-q}^2,\label{idftgn}\\
G^{(1)}(x,x') &=& A^*A'\sum_q\bar{n}_qF_{Q-q}F_{Q'-q},\\
G^{(2)}(x,x') &=& |A|^2|A'|^2\sum_{q\,q'}\bar{n}_q\bar{n}_{q'}\left[F_{Q-q}^2F_{Q'-q'}^2\right.\\ &&\left.+F_{Q-q}F_{Q'-q}F_{Q'-q'}F_{Q-q'}\right],\nonumber\\
  &\approx& |A|^2|A'|^2\left[\sum_{q\,q'}\bar{n}_q\bar{n}_{q'}F_{Q-q}^2F_{Q'-q'}^2\right. \label{G2ft}
  \\ &&\left.+\sum_q\bar{n}_q^2F_{Q-q}^2F_{Q'-q}^2\right].\nonumber
\end{eqnarray}
In this treatment we have assumed that the system is not condensed, a point we address further in Sec.~\ref{SECgftIG}
\subsubsection{High temperature limit}
We now evaluate the  correlation functions (\ref{idftgn})-(\ref{G2ft}) in the high temperature limit, whereby the all quasimomentum states are occupied, however we  assume that no higher bands are occupied (also see \cite{Blakie2004b}). This limit could be arranged by taking a sufficiently deep lattice that $\epsilon_{\rm{BW}}\ll kT\ll\epsilon_{\rm{gap}}$, where $\epsilon_{\rm{BW}}\sim4J$ is the ground band width, and $\epsilon_{\rm{gap}}\sim\hbar\omega_{\rm{Latt}}$ is the excitation gap to higher vibrational bands. In this regime the mean occupation of each of the quasimomentum levels is the same and is given by $\bar{n}_q\approx  N/M$, i.e spatial filling factor $n$. The correlation functions now simplify to
\begin{eqnarray}
 n(x) &=&N|A|^2,\\
G^{(1)}(x,x') &=& \frac{N}{\sqrt{M}}A^*A'F_{Q-Q'},\\
G^{(2)}(x,x') &=& N|A|^2|A'|^2\left[N +\frac{N}{M}F_{Q-Q'}^2\right],\label{G2ht}
\end{eqnarray}
where we have used (\ref{Fsumrule}) in deriving these results.
Comparing these results with Eqs.~(\ref{pMIn})-(\ref{pMIG2}) we see that the measured correlations of the high temperature ideal gas and the Mott-insulator state differ by corrections that are negligible when $N$ is large.

Aspects of the results presented so far have also been considered in Refs.~\cite{Roth2003a,Altman2004a,Bach2004a}. It is of interest to move beyond the pure Bose-Einstein condensate and perfect Mott-insulator approximations  by considering:  (i) the role of interactions in the superfluid state and (ii) tunneling in the Mott-insulator regime. In the following three subsections we turn to developing more realistic descriptions of the optical lattice system that go beyond those simple models and should be widely applicable to experimental regimes.

\subsection{Bogoliubov treatment of the superfluid phase}\label{SECbog}

The Bogoliubov treatment of the superfluid limit of the Bose Hubbard Hamiltonian was given in Refs.~\cite{Oosten2001a,Paraoanu2001a,Rey2003a,Paraoanu2003a,Wild2006a}, and we refer the reader to those references for a detailed description of the method.

Briefly, we recognize that condensation will occur in the 0-quasimomentum mode and  set
\begin{equation}
\hat{b}_0,\hat{b}_0^\dagger\to\sqrt{N_0},\quad N_0\gg1,\label{b0BOG}
\end{equation}
 where $N_0$ is the condensate occupation, and expand the Bose Hubbard Hamiltonian to quadratic order in the remaining quasimomentum operators $\{\hat{b}_{q\ne0}\}$. Performing the canonical transformation
 \begin{equation}
\hat{b}_q=u_q\hat{\beta}_q-v_q\hat{\beta}_{-q}^\dagger, \qquad (q\ne0),\label{bqBOG}
\end{equation}
the Bogoliubov quasiparticle operators $\hat{\beta}_q$ are introduced, where the amplitudes $\{u_q,v_q\}$ are taken to be
\begin{eqnarray}
|u_q|^2 &=&\frac{\epsilon^0_q+n_0U+\epsilon^B_q}{2\epsilon^B_q}, \\
|v_q|^2 &=&\frac{\epsilon^0_q+n_0U-\epsilon^B_q}{2\epsilon^B_q},
\end{eqnarray}
with $n_0=N_0/M$ the average number of condensate atoms per site and
\begin{eqnarray}
\epsilon^0_q &=&4J\sin^2(qa/2),\\
\epsilon^B_q &=& \sqrt{\epsilon^0_q[\epsilon^0_q+2n_0U]}.
\end{eqnarray}
This choice of transformation brings the quadratic  Hamiltonian to diagonal form, i.e. $H\approx{\rm{const.}}\,+\sum_{q\ne0}\epsilon^B_q\hat{\beta}_q^\dagger\hat{\beta}_q$.

Within the Bogoliubov approximation, the $T=0$ state of the system is given by the quasiparticle vacuum state, i.e. $\hat{\beta}_{q\ne0}|\rm{QP}_{\rm{vac}}\rangle=0$, and the condensate population is reduced from the total number of atoms due to the quantum depletion $\tilde{N}$, i.e. $N_0=N-\tilde{N}$ with $\tilde{N}=\sum_{q\ne0}|v_q|^2$. These depleted atoms have an effect on the correlation functions as we now show.

\begin{figure}
 \includegraphics[width=3.3in]{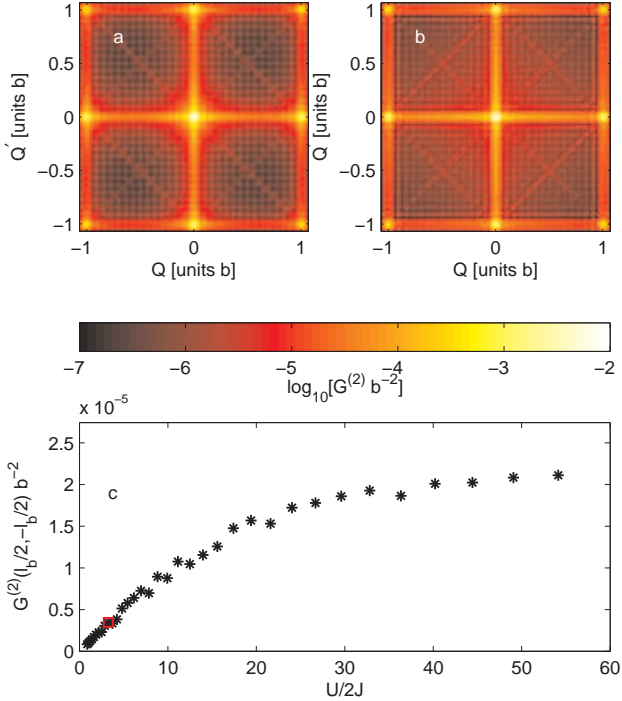}
  \caption{\label{fig:SFcorrG2}
  Second order correlation function, $G^{(2)}$, in the superfluid regime. (a) Bogoliubov result and (b) meanfield decoupling result. (c) Decoupling prediction for the pairing strength, $G^{(2)}(l_b/2,-l_b/2)$ (see text) as $U/J$ is varied. The result for the case shown in (b) is indicated by a square.
 Parameters: Cases (a) and (b) are the same parameters as used in the superfluid  calculations in  Fig.~\ref{fig:IdealDen}. }
  \end{figure}

To calculate the correlation functions we transform the quasi-momentum expansion of the field operators (\ref{psib}) to the quasiparticle form basis. Care needs to be taken with the four-field  correlation function for which we use a number conserving approximation, i.e. the replacement $\hat{b}_0^\dagger\hat{b}_0=N-\sum_{q\ne0}\hat{b}_q^\dagger\hat{b}_q$, rather than simply setting $\hat{b}_0^\dagger\hat{b}_0=N_0$ \footnote{Comparison with  exact many-body solutions shows that the number conserving correction is important, at least in small systems.}. Following this procedure we  obtain
\begin{widetext}
\begin{eqnarray}
{n}(x) &=& |A|^2\big[N_0F_Q^2 +\sum_{q\ne0}v_q^2F_{Q-q}^2\big],\label{BSFn} \\
G^{(1)}(x,x')&=&A^*A'\big[N_0F_QF_{Q'}+\sum_{q\ne0}v_q^2F_{Q-q}F_{Q'-q} \big],\label{BSFG1}\\ 
G^{(2)}(x,x')&=&|A|^2|A'|^2\left\{[N_{0}^{2}+2\sum_{k\ne0}u_{k}^{2}v_{k}^{2}]F_{Q}^{2}F_{Q'}^{2}+\sum_{q\ne0}v_{q}^{2}(N_{0}-2u_{q}^{2})(F_{Q}^{2}F_{Q'-q}^{2}+F_{Q-q}^{2}F_{Q'}^{2})\right.\nonumber\\
&&\left.+\sum_{q\ne0,q'\ne0}v_{q}^{2}v_{q'}^{2}F_{Q-q}^{2}F_{Q'-q'}^{2}+\sum_{q\ne0}v_{q}^{4}F_{Q-q}^{2}F_{Q'-q}^{2}+\sum_{q\ne0}u_{q}^{2}v_{q}^{2}F_{Q-q}^{2}F_{Q'+q}^{2}\right\}.\label{G2Bog}
 \end{eqnarray}
 In Figs.~\ref{fig:IdealDen}(b) and \ref{fig:SFcorrG2}(a)  we show results for $n(x)$ and $G^{(2)}(x,x')$ in the Bogoliubov approximation.

 For this case it is interesting to consider the covariance which, according to Eq.~(\ref{CovReln}), is
 \begin{eqnarray}
 C(x,x')&=&|A|^2|A'|^2\left\{2\sum_{k\ne0}u_{k}^{2}v_{k}^{2}F_{Q}^{2}F_{Q'}^{2}+\sum_{q\ne0}\left[v_{q}^{4} F_{Q-q}^{2} F_{Q'-q}^{2}  + u_{q}^{2} v_{q}^{2} F_{Q-q}^{2}F_{Q'+q}^{2}\right] -2\sum_{q\ne0}u_{q}^{2}v_{q}^{2}\left(F_{Q}^{2}F_{Q'-q}^{2}+F_{Q-q}^{2}F_{Q'}^{2}\right)\right\}\nonumber\\&&-|A|^2(N_{0}F_{Q}^{2}+\sum_{q\ne0}v_{q}^{2}F_{Q-q}^{2})\delta(x-x').\label{BogSFC}
 \end{eqnarray}
 \end{widetext}
 We note three observable effects of the quantum depletion in Eq.~(\ref{BogSFC}). (i)  The $v_q^4$-term represents the additional fluctuations due to the quantum depletion which is of the same form as the thermal depletion for the finite temperature ideal gas. (ii) The $\sum_{q\ne0}u_{q}^{2}v_{q}^{2}F_{Q-q}^{2}F_{Q'+q}^{2}$-term represents the pairing between particles in the quantum depletion which, due to momentum conservations, appears as a correlations between density fluctuations at $+q$ and $-q$. (iii) The $-2\sum_{q\ne0}u_{q}^{2}v_{q}^{2}\left(F_{Q}^{2}F_{Q'-q}^{2}+F_{Q-q}^{2}F_{Q'}^{2}\right)$-term represents the anti-correlation between depleted atoms and atoms in the condensate. This last term is absent if we do not make the number-conserving approximation.

 In Fig.~\ref{fig:SFcorrG2}(a) the depletion and pairing effects can been seen. The weak diagonal line (e.g. along $Q=Q'$) arises from the quantum depletion, whereas the other diagonal (e.g. $Q=-Q'$) arises from the pairing in the system. This figure shows that the pairing and depletion are approximately the same size, and are much smaller than the signal due to the condensate which dominates at $Q=0$ or $Q'=0$. Normalizing the correlation function to obtain $g^{(2)}(x,x')$ significantly enhances the relative size of the pairing and depletion. 
 
 \begin{figure}
 \includegraphics[width=3.3in]{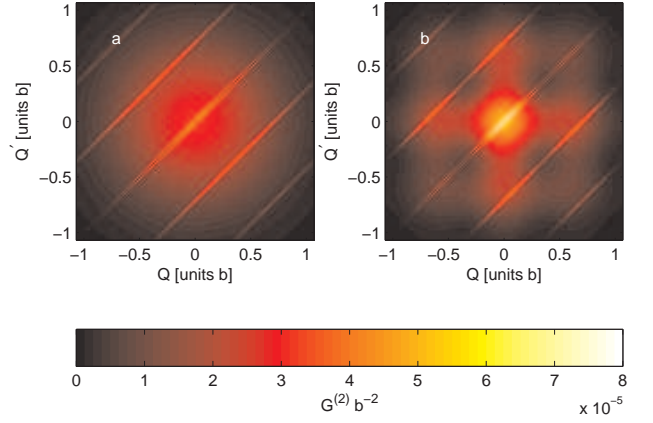}
  \caption{\label{fig:PHMott}
  Second order correlation functions in the Mott-insulator regime. (a) Perfect Mott-insulator state and (b) particle-hole corrections to the perfect Mott-insulator state.
 The same parameters are used as for the Mott-insulator calculations in  Fig.~\ref{fig:IdealDen}. 
 }
  \end{figure}

\subsection{Decoupling approach}\label{SECDecoupling}
The meanfield decoupling approach is widely used to describe the many-body state of the system in both the Mott-insulator and the superfluid regions.
The essence of this approach   \cite{Sheshadri1995} is the approximation $\hat{a}_j^\dagger \hat{a}_{k} \approx \hat{a}_j^\dagger \alpha_{k}+ \alpha_j^* \hat{a}_{k} - \alpha_j^* \alpha_{k}$, with $\alpha_j = \langle \hat{a}_j\rangle$.  This turns the grand canonical form of  Bose Hubbard Hamiltonian (\ref{BHH})  into   the \emph{site-decoupled} form \footnote{Coupling between sites still remains through the mean field parameters $\alpha_j$.}, i.e.
\begin{equation}
\hat{H}_{\rm{BH}}-\mu \hat{N}  \to \hat{H}_{\rm{D}}=\sum_{j=1}^M\hat{H}_{j},\label{HD}
\end{equation}
where $\hat{N}=\sum_j\hat{a}_j^\dagger\hat{a}_j$ is the number operator, $\mu$ is the chemical potential \footnote{We introduce $\mu$ into this formalism to fix the mean number of particles per site.} and
\begin{eqnarray}
\hat{H}_j&=&\left(\frac{U}{2}\hat{a}_j^\dagger\hat{a}_j^\dagger\hat{a}_j\hat{a}_j-J\left[\hat{a}_j^\dagger(\alpha_{j-1}+\alpha_{j+1})+ \rm{h.c}\right]  \right.  \\
&&\left.+J\alpha^*_j(\alpha_{j-1}+\alpha_{j+1})-\mu\hat{a}_j^\dagger\hat{a}_j\right).\nonumber
\end{eqnarray}
The manybody  ground state of  (\ref{HD}) can be written in the (decoupled) product form
\begin{equation}
|\Phi_{D}\rangle = \prod_{j=0}^M|\psi_j\rangle,\label{psisinglesite}
\end{equation}
where the $|\psi_j\rangle$  is the ground state of $\hat{H}_j$, and can be convenient expressed in terms of the number states of $\hat{a}_j^\dagger\hat{a}_j$.
In the translationally invariant system we consider here, each site is equivalent and we can drop the $j$-label (i.e. setting $\alpha_j\to \alpha$) and solve the single site problem $\hat{H}_j$. 

The advantage of this approach is that it is able to represent number states and coherent states, and thus contains as limits the pure Bose-Einstein condensate and perfect Mott-insulator states. However, the decoupling approximation neglects correlations between sites beyond those described by the coherent coupling $\alpha$, and thus can only provide a rather limited description of spatial correlations in the system.
\begin{figure}
 \includegraphics[width=3.2in]{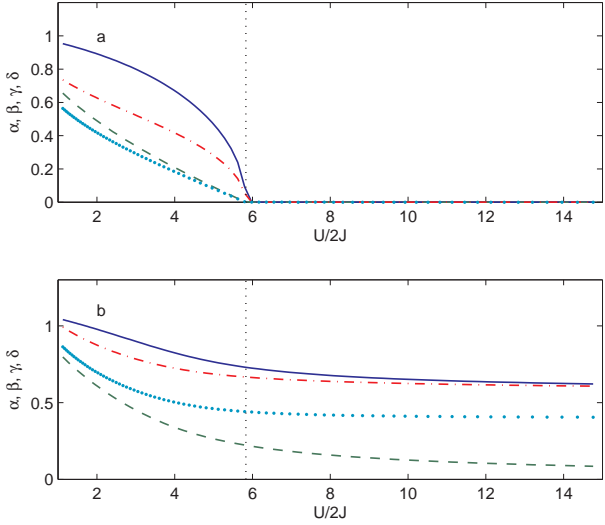}
  \caption{\label{fig:alpha}
  Behavior of meanfield decoupling parameters as lattice depth changes. (a) Commensurately filled lattice with $n=1$ and (b) incommensurate case with $n=1.2$.  Results for  $\alpha$ (solid line), $\beta$ (dashed line), $\gamma$ (dash-dot line) and $\delta$  (dotted line) are shown. Meanfield estimate of the Mott-insulator transition for $n=1$ is indicated with the vertical dotted line.}
  \end{figure}

Within the decoupling approximation we obtain the following expressions for the correlation functions
\begin{eqnarray}
n(x) &=& |A|^2\left[(N-M|\alpha|^2)+M|\alpha|^2F_Q^2\right],\label{dcn}\\
G^{(1)}(x,x') &=& A^*\!A'\left[\frac{N}{\sqrt{M}}\frac{\chi_C}{n}F_{Q-Q'} \!
  +M|\alpha|^2F_QF_{Q'}\right],\label{dcG1}\\
G^{(2)}(x,x') &=& |A|^2|A'|^2[M\chi_A\!+M^2|\alpha|^4F_Q^2F_{Q'}^2 \label{dcG2}\\
&&+M (\chi_D ^2F^2_{Q+Q'}+ \chi_C ^2F^2_{Q-Q'})+M\chi_B(F_Q^2+F_{Q'}^2)\nonumber\\
&&+2M^{\frac{3}{2}}F_QF_{Q'}(|\alpha|^2\chi_CF_{Q-Q'}+\Re\{\alpha^2\chi_D^*\}F_{Q+Q'})]\nonumber,
\end{eqnarray}
where $\beta\equiv\langle\hat{a}_j\hat{a}_j\rangle$, $\gamma\equiv\langle\hat{a}_j^\dagger\hat{a}_j\hat{a}_j\rangle$, $\delta\equiv\langle\hat{a}_j^\dagger\hat{a}_j^\dagger\hat{a}_j\hat{a}_j\rangle$, are parameters that can be determined from the meanfield solution (i.e. $|\psi_j\rangle$ in Eq.~(\ref{psisinglesite})) and
\begin{eqnarray}
\chi_A&=&M\chi_C^2+\delta-|\beta|^2+2|\alpha|^2(4n-3|\alpha|^2)\nonumber\\ &&+4\Re\{\alpha(\alpha\beta^*-\gamma^*)\} -2n^2,\label{EqChiA}\\
\chi_B &=&(M-4)(|\alpha|^2\chi_C-2\Re\{\alpha(\alpha\beta^*-\gamma^*)\},\label{EqChiB}\\
\chi_C &=&n-|\alpha|^2,\label{EqChiC}\\
\chi_D &=&\beta-\alpha^2,\label{EqChiD}
\end{eqnarray} 
where $\Re$ means the real part.
In Fig.~\ref{fig:alpha} we show the typical behavior of   $\{\alpha,\beta,\gamma,\delta\}$ parameters as the lattice depth increases. For the commensurate case [Fig.~\ref{fig:alpha}(a)] we see that all of these parameters go to zero when the Mott-insulator transition occurs. I.e. when $(U/J)>g_c$, the local solutions, $|\psi_j\rangle$,  approach number states (if $n$ is an integer). In the $U/J\to0$ limit, the local solutions $|\psi_j\rangle$ approach coherent states with $\alpha\to\sqrt{n}$, $\beta\to n$, $\gamma\to n^{3/2}$ and $\delta\to n^2$. For the incommensurately filled lattice [Fig.~\ref{fig:alpha}(b)], the Mott-insulator transition does not occur and residual coherence exists in the deep lattice limit.

The above results for $\{\alpha,\beta,\gamma,\delta\}$  allow us to understand how the parameters $\{\chi_A,\chi_B,\chi_C,\chi_D\}$ behave in the limiting cases:
\paragraph{Superfluid limit:}
 As $U\to0$,  we have $\chi_A,\chi_B,\chi_C,\chi_D\to0$ and $G^{(1)}$ and $G^{(2)}$ agree with the expressions of the pure Bose-Einstein condensate. (Note that in this limit the lower two lines of Eq.~(\ref{dcG2})  vanish).
 In Figs.~\ref{fig:IdealDen}(b) and \ref{fig:SFcorrG2}(b)  we show decoupling results for $n(x)$ and $G^{(2)}(x,x')$ in the superfluid limit. 

\paragraph{Mott-insulator limit:}
For $(U/J)>g_c$ (and $n$=integer) we have that $\chi_A\to Nn-n(n+1)$ and $\chi_C\to n$, while $\chi_B,\chi_D\to0$. Under these conditions $G^{(1)}$ and $G^{(2)}$ agree with the expressions of the  perfect Mott state. (Again we note that in this limit the lower two lines of Eq.~(\ref{dcG2})  vanish).

\subsubsection{Comparison of decoupling and Bogoliubov approaches}
It is of interest to compare the Bogoliubov and decoupling results in the superfluid regime to better understand the physics captured by each approximation (see Figs.~\ref{fig:SFcorrG2}(a) and (b)).
As the value of $U/J$ increases both methods predict an increase in the quantum depletion, however for the Bogoliubov approach the depletion is colored according to $v_q^4$ whereas for the Decoupling approach the depletion is uniform in quasi-momentum space and proportional to $(n-|\alpha|^2)$, i.e. $\chi_C$.  These differences are most apparent in the density predictions for each method shown in Fig.~\ref{fig:IdealDen}(b).

While the Bogoliubov approach predicts no transition, and is hence invalid in the deep lattice limit, at a finite value of $U/J$ the parameter $\alpha$ goes to zero and a purely incoherent interference pattern is produced. 
Similarly both methods predict pairing, colored and proportional to $u_q^2v_q^2$ for Bogoliubov, while the decoupling result is uniform and proportional to $(\beta-\alpha^2)$, i.e. $\chi_D$. The uniform nature of the depletion and pairing predicted by the decoupling approach  indicates one of the main deficiencies of this theory: the decoupling approximation makes all single particle excitations of the system  (other than the  $q=0$ condensate) degenerate. 

\subsubsection{Strength of pairing correlations}
In view of the reasonable comparison of the decoupling and Bogoliubov approaches, in Fig.~\ref{fig:SFcorrG2}(c) we show the strength of the pairing  as predicted by the decoupling approximation. To do this plot the value of $G^{(2)}$ at the location $x=l_b/2$ and $x'=\!-\!l_b/2$ (i.e. $Q=b/2$ and $Q'=\!-\!b/2$). As the lattice depth increases  the value of the pairing is observed to increase, saturating  to a maximum value after the system passes through the Mott-insulator transition.

\subsection{Particle-hole corrections to the Mott-insulator phase}\label{SECphMI}
Recent experiments done in 3D and 2D systems \cite{Gerbier2005,Spielman2007a} have observed
that phase coherence on short length
scales persists even deep in the insulating phase.
This behavior can be attributed to a coherent admixture of
particle-hole pairs to the perfect Mott state for small but finite tunneling and can be estimated by using first order perturbation theory.
We consider first  a homogeneous system with filling factor $n$.
In the limit of infinitely strong repulsion, $U/J\rightarrow\infty$, the
ground state is what we called a perfect Mott-insulator, Eq.~(\ref{pmi}).
As shown in Fig.~\ref{fig:IdealDen}(c) the expanded density distribution of this state  is flat and consequently has zero fringe visibility.
By treating  the tunneling term as a perturbation to the interaction term, one can account for finite tunneling corrections
and to first order in $J/U$ one gets that on top of the Mott-insulator core the system has an admixture of nearest-neighbor ``particle-hole'' excitations (a site with an additional particle while one of its  nearest neighbors is missing a particle):
\begin{equation}
|\Phi_{\rm{M1}}\rangle \approx |\Phi_{\rm{M}}\rangle+\frac{J}{U}\sum_{\langle i,j \rangle }\hat{a}_i^\dagger  \hat{a}_j|\Phi_{\rm{M}}\rangle,
\end{equation} The restoration of the short range phase coherence induced by such excitations  is signaled  in  the correlation functions which become:
\begin{eqnarray}
 n_{\rm{M1}} &=& n_{\rm{M}}-\frac{A^2 2 N(n+1)}{U} E(Q), \label{phMIn}\\
G_{\rm{M1}}^{(1)} &=& G_{\rm{M}}^{(1)} -\frac{ A^*A' N(n+1)}{\sqrt{M} U}F_{Q-Q'}[E(Q)+E(Q')], \label{phMIG1}\\
G_{\rm{M1}}^{(2)} &=&G_{\rm{M}}^{(2)}-\left(|A|^2|A'|^2 \frac{2 M N [E(Q)+E(Q')] }{U} \right)\times \label{phMIG2}\\
 &&\left[ n(n+1)(1+ \frac{F_{Q-Q'}^2}{M}) -\frac{1}{M}(2n^2+3n+1)\right],\nonumber
  \end{eqnarray}
  where  $E(Q)=-2J\cos(Qa)$  
  is the single particle dispersion relation.   In Fig.~\ref{fig:IdealDen}(c) the density $ n_{\rm{M1}}(x)$ is shown while in  Fig.~\ref{fig:PHMott}(a) and (b) the second order correlation functions for the perfect Mott-insulator and the case including particle-holes corrections are compared. A rather significant effect of these particle-hole correlations are observed as long wavelength modulations of the density, characterized by the term $E(Q)$ given above.
  
  While the particle-hole modulation [see Fig.~\ref{fig:IdealDen}(c)]  to the expanded density distribution has been experimentally measured \cite{Gerbier2005,Spielman2007a} the corrections to the noise correlations has not yet been observed. 
   One possible reason is that in the noise correlation experiments part of the superfluid component had to be masked out before computing the correlation functions.
   
 Even though here we will  limit  our analysis to first order corrections we conclude this subsection 
  by briefly outlining  the basic physics beyond it:  
  Recent theoretical analysis based on the Random Phase Approximation (RPA) \cite{Gerbierb2005} supports a physical picture of the system as
a (dilute) gas of partice-hole pairs, mobile through the lattice,
on top of a regularly filled Mott-insulator. Consequently what happens   for larger  $J/U$ values
is the development of  phase coherence at longer length
scales as the particle-hole pair size extends over larger distance scales than a single lattice constant  through higher-order tunneling
events  which become non-negligible.

\section{Gaussian approach (1D)}\label{SEC:GRESULTs}
In the previous section results for the expanded correlations of a one-dimensional system of bosons in a translationally invariant lattice were derived. Extending these results, particularly for $G^{(2)}$, to higher dimensions and inhomogeneous potentials is difficult. Here we develop a different approach:   we approximate the expanded matter wave field as being Gaussian, so that all higher order correlation functions can be determined from the low order moments, i.e.~in terms of the condensate wavefunction, and the normal and anomalous density matrices (introduced below). 
 This approach avoids directly evaluating the four-point correlation function and is thus better suited to numerical calculations, which are generally required for inhomogeneous situations (which we consider in the next section).
 We emphasize that while a Gaussian approach is in general insufficient for capturing the \emph{in situ} many-body physics of the lattice system, particularly in the Mott-insulator regime, we show that it provides a good representation of the low order noise correlations of the expanded system for the various approaches we have considered so far.

The general Gaussian expression for the $G^{(2)}(x,x')$ correlation function is
\begin{eqnarray}
G^{(2)}(x,x') &\approx& |G^{(1)}(x,x')|^2+n(x)n(x')-|\Phi_0(x)|^2|\Phi_0(x')|^2\nonumber\\
&&+\left\{\Phi_0^*(x)\Phi^*_0(x')M(x,x')+\rm{c.c}\right\}\label{G2Gaussian}
\\&&+|M(x,x')|^2,\nonumber
\end{eqnarray}
where we have introduced the following quantities:

{\bf Condensate orbital:} $\Phi_0(x)$. When the system is in the superfluid phase this quantity is non-zero and reflects the long-range order in the system. The condensate orbital can be obtained as an expectation of the field operator $\Phi_0(x)=\langle \hat{\psi}_g(x)\rangle$ in symmetry broken approaches or by the Penrose-Onsager criterion \cite{Penrose1956a}.

{\bf Anomalous average:}  $M(x,x')$  (or anomalous density matrix). In symmetry broken approaches, when a condensate exists in the system a fluctuation operator can be defined as
\begin{equation}
\hat{\xi}(x) = \hat{\psi}_g(x)-\Phi_0(x),
\end{equation}
such that $\langle \hat{\xi}(x)\rangle=0$. The anomalous average is then given as
\begin{equation}
M(x,x')=\left\langle\hat{\xi}(x)\hat{\xi}(x')\right\rangle.\label{Mdefn}
\end{equation}

The first line of Eq.~(\ref{G2Gaussian}) was used in Ref.~\cite{Naraschewski1999a} to analyze  the \emph{in situ} correlations of a harmonically trapped Bose gas. Here we use the more general expression which also 
includes anomalous terms (e.g. see Chapter 4 of Ref.~\cite{blaizot-ripka}).
We reiterate that here we are not approximating the \emph{in situ} Bose-Hubbard state as being Gaussian, only the matter wave field after expansion.
Indeed, of the various approaches we have considered, only the pure superfluid, ideal gas  
and Bogoliubov methods are themselves Gaussian descriptions of the many- 
body state. The perfect Mott insulator,  decoupling and particle-hole approaches are non-gaussian states and for  these cases it  
is not clear that expressing the $G^{(2)}$ function in terms of lower  
order moments (\ref{G2Gaussian}) is appropriate -- an issue we examine  
more closely in the remainder of this section.

More generally, the expanded correlation functions can be expressed in  
terms of cumulants which, as opposed to operator moments, tend to  
become smaller at high orders \cite{Koehler2002a}. Indeed, for a  
system described by a Gaussian density matrix, all cumulants higher  
than second order (which we refer to here as first order correlation  
functions) are zero \footnote{In terms of the notation of Ref.~ 
\cite{Koehler2002a}, our condensate orbital is the first order  
cumulant, $M(x,x')$ corresponds to the pair function (a second order  
cumulant), and $G^{(1)}(x,x')$ corresponds to the full one-body  
density matrix, with the non-condensate density matrix $G^{(1)}(x,x')- 
\Phi_0(x)\Phi_0^*(x')$ corresponding to the other second order  
cumulant.}. This approach would allow one to systematically go beyond  
the Gaussian approach by considering higher order cumulants, however  
our results here suggest this is unnecessary. We also note that a more precise number conserving definition of the  
anomalous average could be adopted \cite{Morgan2004a}, however for the  
purposes of this paper the  definition given in (\ref{Mdefn}) is  
adequate.

 We now apply the Gaussian approach to the 1D translationally invariant cases considered in the last section to justify its use.
\subsection{Gaussian pure Bose-Einstein condensate}
 For the pure Bose-Einstein condensate state considered in Sec.~\ref{SECidSF} we find that $\Phi_0(x)=A\sqrt{N}F_Q$ and $M(x,x')=0$. So that from Eq.~(\ref{G2Gaussian}) we obtain
 \begin{equation}
G^{(2)}(x,x')=N^2|A|^2|A'|^2 F_Q^2F_{Q'}^2,
 \end{equation}
 which differs from our earlier result (\ref{iSFG2}) only by the $N(N-1)$ coefficient on $G^{(2)}$ which is negligible when $N$ is large. More generally, in this limit the system is completely described by the coherent condensate mode and all higher order correlation functions trivially factorize.

\subsection{Gaussian perfect Mott-Insulator}
For the perfect Mott-insulator state  considered in Sec.~\ref{SECpMI} the condensate and anomalous average are both zero and  from Eqs.~(\ref{pMIn}), (\ref{pMIG1}) and  (\ref{G2Gaussian}) we obtain
\begin{equation}
G^{(2)}(x,x') = N|A|^2|A'|^2\left[N +\frac{N}{M}F_{Q-Q'}^2\right].
\end{equation}
This differs from the exact  result (\ref{pMIG2}) by a correction to the first $N$ in the square brackets that is negligible for large $N$.

\subsection{Gaussian finite temperature ideal gas}\label{SECgftIG}
 For the finite temperature ideal gas  considered in Sec.~\ref{SECftIG} the Gaussian approach gives identical results to those in Eq.~(\ref{G2ft}) [hence we also obtain the same predictions for the high temperature limit given in Eq.~(\ref{G2ht})].
 We note that in obtaining these results we assumed the condensate orbital and the anomalous average are zero. The no-condensate assumption was also assumed in the derivation presented in Sec.~\ref{SECftIG}, whereby all the modes were taken to fluctuate according to the grand canonical ensemble. When a condensate is present the $\Phi_0(x)$ corrections in Eq.~(\ref{G2Gaussian}) are important (this issue is discussed more fully in Ref.~\cite{Naraschewski1999a}). In this case the $q=0$ mode defines the condensate, i.e. $\Phi_0(x)=A\sqrt{\bar{n}_0}F_Q$, and in the ideal limit (where there is no anomalous correlation)   we obtain
 \begin{widetext}
\begin{eqnarray} 
G^{(2)}(x,x') &=& |A|^2|A'|^2\left\{\sum_{q\,q'}\bar{n}_q\bar{n}_{q'}\left[F_{Q-q}^2F_{Q'-q'}^2 +F_{Q-q}F_{Q'-q}F_{Q'-q'}F_{Q-q'}\right]-\bar{n}_0^2F_Q^2F_{Q'}^2\right\}, \\
  &\approx& |A|^2|A'|^2\left[\sum_{q\,q'}\bar{n}_q\bar{n}_{q'}F_{Q-q}^2F_{Q'-q'}^2 +\sum_{q\ne0}\bar{n}_q^2F_{Q-q}^2F_{Q'-q}^2\right]. \label{gG2ft}
\end{eqnarray}

\subsection{Gaussian Bogoliubov treatment of the superfluid phase}
 The Bogoliubov treatment given in Sec.~\ref{SECbog} is essentially a Gaussian theory and is well suited to the development of this section. The condensate orbital is  $\Phi_0(x)=A\sqrt{N_0}F_Q$, so that the fluctuation operator is given by $\hat{\xi}(x)=A\sum_{q\ne0}(u_q\hat{\beta}_q-v_q\hat{\beta}^\dagger_{-q})F_{Q-q}$ and the anomalous average is
 \begin{equation}
 M(x,x')=-AA'\sum_{q\ne0}u_qv_qF_{Q-q}F_{Q'+q}.
 \end{equation}
 Thus using these results and those in Eqs.~(\ref{BSFn}), (\ref{BSFG1}) and (\ref{G2Gaussian}), we obtain
 \begin{eqnarray} 
G^{(2)}(x,x')&=&|A|^2|A'|^2\left[N_{0}^{2}F_{Q}^{2}F_{Q'}^{2}\ +\sum_{q\ne0}v_{q}^{2}N_{0}(F_{Q}^{2}F_{Q'-q}^{2}+F_{Q-q}^{2}F_{Q'}^{2})\right.\nonumber\\
&&\left.+\sum_{q\,q'\ne0}v_{q}^{2}v_{q'}^{2}F_{Q-q}^{2}F_{Q'-q'}^{2}+\sum_{q\ne0}v_{q}^{4}F_{Q-q}^{2}F_{Q'-q}^{2} +\sum_{q\ne0}u_{q}^{2}v_{q}^{2}F_{Q-q}^{2}F_{Q'+q}^{2}\right].
 \end{eqnarray}
 Compared to Eq.~(\ref{G2Bog}) this result lacks the anti-correlation between the condensate and quasi-particle modes, but otherwise shows the same behavior for the quantum depletion and pairing. The pairing term arises from the anomalous average, and shows the importance of including the anomalous average in any description correlations for the interacting superfluid regime.
 The anti-correlation effects could be recovered by using a number-conserving generalization to the definition of the anomalous average, but we will not address this further here.

\subsection{Gaussian form of particle-hole corrections to the Mott-insulator phase}
Using  the Gaussian approximation,  one can  calculate the particle-hole corrections to the perfect Mott-Insulator state  considered in Sec.~\ref{SECphMI} by  using  Eq.~(\ref{G2Gaussian}) with the results given in Eqs.~(\ref{phMIn}) and (\ref{phMIG1}), and  by setting both the condensate and anomalous average to  zero. This procedure yields, to first order in $J/U$, the following expressions for the correlations functions:
\begin{eqnarray}
G^{(2)}(x,x') &=& |G_{\rm{M}}^{(1)}(x,x')|^2+n_{\rm{M}}(x)n_{\rm{M1}}(x')+2{\Re}[G_{\rm{M}}^{(1)}(x,x') G_{\rm{M1}}^{(1)}(x,x')]+n_{\rm{M}}(x)n_{\rm{M1}}(x')+n_{\rm{M1}}(x)n_{\rm{M1}}(x'),\\
 &=&N|A|^2|A'|^2\left[N +\frac{N}{M}F_{Q-Q'}^2\right]
-\left(|A|^2|A'|^2 \frac{2 M N [E(Q)+E(Q')] }{U} \right)\left[ n(n+1)\left(1+ \frac{F_{Q-Q'}^2}{M}\right) \right].\label{G2phpert2}
\end{eqnarray}
Similarly to the perfect Mott-insulator situation, the Gaussian approximation results differ from the exact  results, Eq.~(\ref{phMIG2}), by a term  in the last square brackets of Eq.~(\ref{G2phpert2})  that is negligible in the thermodynamic limit.

\subsection{Gaussian decoupling approach}
The mean-field decoupling approach was introduced in Sec.~\ref{SECDecoupling}. The condensate orbital is given by $\Phi_0(x)=\langle\hat{\psi}_g(x)\rangle=\sqrt{M}\alpha AF_Q$, so that the fluctuation operator is given by $\hat{\xi}(x)=\hat{\psi}_g(x)-\Phi_0(x)$ and
\begin{equation}
M(x,x')=AA'\sqrt{M}(\beta-\alpha^2) F_{Q+Q'}.
\end{equation}
Using this result and  Eqs.~(\ref{dcn}), (\ref{dcG1}) and (\ref{G2Gaussian}) we obtain
 \begin{eqnarray} 
 G^{(2)}(x,x') &=& |A|^2|A'|^2\left[M\tilde{\chi}_A+M^2|\alpha|^4F_Q^2F_{Q'}^2
+M (\chi_D ^2F^2_{Q+Q'} +\chi_C ^2F^2_{Q-Q'})\right.\nonumber\\ &&\left.+M\tilde{\chi}_B(F_Q^2+F_{Q'}^2)+2M^{\frac{3}{2}}F_QF_{Q'}\left(|\alpha|^2\chi_CF_{Q-Q'}+\Re\{\alpha^2\chi_D^*\}F_{Q+Q'}\right)\right],\label{G2decouplingGaus}
 \end{eqnarray}
 \end{widetext}
with ${\chi}_C$ and $\chi_D$ as defined in Eqs.~(\ref{EqChiC}) and (\ref{EqChiD}), and
\begin{eqnarray}
\tilde{\chi}_A &=& M\chi_C^2,\label{EqtChiA}\\
\tilde{\chi}_B &=& M|\alpha|^2\chi_C.\label{EqtChiB}
\end{eqnarray}

   \begin{figure}
 \includegraphics[width=3.2in]{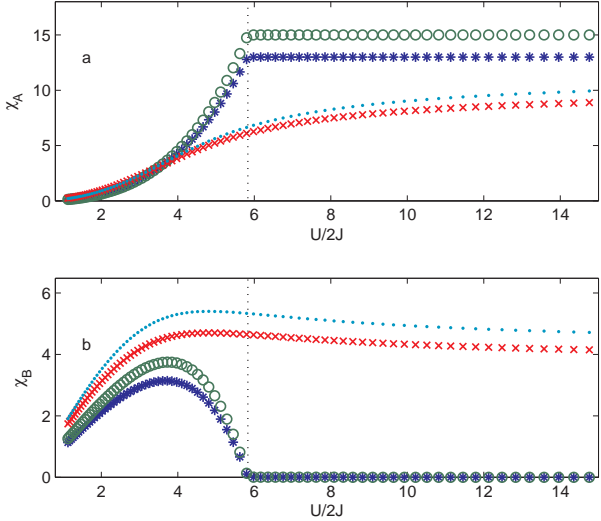}
  \caption{\label{fig:DecouplGausVsFull}
  Comparison of Gaussian and full decoupling results. (a) Comparison of $\chi_A$ for $n=1$ (stars) and $n=1.2$ (crosses), and $\tilde{\chi}_A$ for $n=1$ (circles) and $n=1.2$ (dots).
  (b) Comparison of $\chi_B$ for $n=1$ (stars) and $n=1.2$ (crosses), and $\tilde{\chi}_B$ for $n=1$ (circles) and $n=1.2$ (dots).
Meanfield estimate of Mott-insulator transition  for $n=1$ is indicated with the vertical dotted line.}
  \end{figure}

\subsubsection{Comparison with full decoupling result}
Here we compare the differences between the Gaussian decoupling result above with the full result developed in Sec.~\ref{SECDecoupling}. As emphasized in our choice of notation above, these differences arise between $\chi_A$ [Eq.~(\ref{EqChiA})] and $\chi_B$ [Eq.~(\ref{EqChiB})] and the respective quantities $\tilde{\chi}_A$ [Eq.~(\ref{EqtChiA})] and $\tilde{\chi}_B$ [Eq.~(\ref{EqtChiB})]. These quantities are compared for cases of commensurate and incommensurate filled lattices in Fig.~\ref{fig:DecouplGausVsFull}.

In Fig.~\ref{fig:DecouplGausVsFull}(a) $\chi_A$ and $\tilde{\chi}_A$ are considered. In the pure Bose-Einstein condensate limit  (i.e. small $U/2J$) both results agree with  $\chi_A\to0$ and $\tilde{\chi}_A\to0$. The deep lattice (i.e. large $U/2J$) behavior depends on whether the system is commensurately filled or not. For $n$=integer the Mott-insulator transition occurs and beyond this we have $\chi_A\to n(N-n-1)$ and $\tilde{\chi}_A\to nN$ which agree to order $n/N$. For the incommensurate case no sharp transition is observed, however $\chi_A$ and $\tilde{\chi}_A$ are seen to be in reasonable agreement for all values of the lattice depth. 

Second, we compare $\chi_B$ and $\tilde{\chi}_B$ in Fig.~\ref{fig:DecouplGausVsFull}(b).   For the commensurate lattice both quantities approach the value of 0 in the pure Bose-Einstein condensate and in the Mott-insulator limits. In between those cases, in the interacting superfluid regime, $\chi_B$ and $\tilde{\chi}_B$ take non-zero values and are similarly behaved. For the incommensurate case $\chi_B$ and $\tilde{\chi}_B$ have appreciable values in the deep lattice limit.

 These comparisons give us confidence that the Gaussian decoupling  
approach provides a reasonably accurate representation of the full  
decoupling results. We also note that in general the agreement  
improves as the filling factor increases.

More generally, the results of this section strongly support that a  
Gaussian representation of the expanded correlation functions is  
accurate even for non-Gaussian states many-body states, such as the  
perfect Mott insulator, decoupling state and particle-hole result. For  
example, in the decoupling approach the single site moments, $\alpha, 
\beta,\gamma$ and $\delta$ are all important and poorly approximated  
as Gaussians (i.e. approximating $\gamma$ and $\delta$ in terms of  
products of $\alpha$ and $\beta$). 
However, the far-field matter wave results from the interference of atoms from all parts of the original lattice. In this case the central limit theorem ensures the measured statistics (in momentum space) are closer to Gaussian than were those of the \emph{in situ} state. This suggests that the general applicability and accuracy of the Gaussian approach should be quite broad if we can accurately calculate the low order moments.

\section{Inhomogeneous Gaussian theory (3D)}\label{SEC:3DRESULTs}
  It would be feasible to extend the exact evaluations of $G^{(2)}$ considered in  
Sec.~\ref{SEC:RESULTs} to the 3D case for a translationally invariant  
lattice. However, in the presence of additional inhomogeneous  
potentials (as in experiments), the generic exact results for the  
second order correlation function would involve computing the four- 
point correlation functions  $\left(\langle \hat{a}_i^\dagger\hat{a}_j^ 
\dagger\hat{a}_k\hat{a}_l\rangle\right)$ without the simplifications afforded by translational invariance. In the context of 3D calculations, such results would  
be computationally impractical to deal with.

  The primary advantage of the gaussian approach we have justified in  
the previous section is that it offers a convenient method for  
extending the various techniques for calculating
correlations to inhomogeneous 3D situations, and only involves 2-point  
correlation functions. This ultimately lends itself to a numerically  
tractable theory for calculating results that can be compared with  
experiments.

  In this section we consider a full 3D formulation, and thus extend  
our previous abbreviated notation for this purpose. In particular, in  
this section we will use the following: 
 \begin{eqnarray}
 A(\mathbf{x},t)&\to& A ,\nonumber\\
 A(\mathbf{x}',t)&\to& A',\nonumber\\
 F(\mathbf{Q}(\mathbf{x})-\mathbf{q})&\to& F_{\mathbf{Q}-\mathbf{q}},  \end{eqnarray} and so on.
 
For application to the inhomogeneous system it is useful to define the momentum space representation of system quantities. In this section we will indicate Fourier transformed quantities with a tilde, e.g. the discrete Fourier transform of the density yields  
\begin{equation}
\tilde{n}_{\mathbf{q}} = \frac{1}{M^{\frac{3}{2}}}\sum_{\mathbf{j}}n_{\mathbf{j}}e^{i\mathbf{q}\cdot\mathbf{R_j}},
\end{equation} 
where $\mathbf{q}$ is limited to the discrete set of allowed quasi-momentum values.

 It is also convenient to extend this definition to the continuous Fourier transform which we can parameterize with far-field momentum $\mathbf{Q}$. We can easily show that 
\begin{equation}
\tilde{n}_{\mathbf{Q}}=\frac{1}{M^{\frac{3}{2}}}\sum_{\mathbf{j}}n_{\mathbf{j}}e^{i\mathbf{Q}\cdot \mathbf{R_j}}=\frac{1}{M^{\frac{3}{2}}}\sum_{\mathbf{q}}\tilde{n}_{\mathbf{q}}F_{\mathbf{Q}-\mathbf{q}},
\end{equation}
i.e. the Fourier transformed density (indicated with a subscript $\mathbf{Q}$) contains the $F_{\mathbf{Q}-\mathbf{q}}$ implicitly, and thus using this form affords us a more compact notation. We emphasize that this quantity is not the momentum density, which could be obtained from the Fourier transform of the field operator.
In many of the derivations that follow we also make use of the identity
\begin{equation}
\tilde{n}_{\mathbf{Q}-\mathbf{Q'}}=\frac{1}{M^3}\sum_{\mathbf{q}\,\mathbf{q'}}\tilde{n}_{\mathbf{q}-\mathbf{q'}}F_{\mathbf{Q}-\mathbf{q}}F_{\mathbf{Q'}-\mathbf{q'}}.
\end{equation}

\subsection{Inhomogeneous pure Bose-Einstein condensate}
The pure Bose-Einstein condensate in the inhomogeneous system is described by the discrete Gross-Pitaevskii equation
\begin{equation}
\mu z_{{\mathbf{i}}}=-J\sum_{\langle {\mathbf{i}},{\mathbf{j}}\rangle}z_{{\mathbf{j}}}+\epsilon_{{\mathbf{i}}}z_{{\mathbf{i}}}+U|z_{{\mathbf{i}}}|^2z_{{\mathbf{i}}},
\end{equation}
where $z_{{\mathbf{i}}}=\langle \hat{a}_{{\mathbf{i}}}\rangle$ is the condensate amplitude on site ${\mathbf{i}}$, and $\mu$ is the chemical potential chosen to ensure $\sum_{{\mathbf{i}}}|z_{{\mathbf{i}}}|^2=N$.
Applying the Gaussian formalism requires the following quantities
\begin{eqnarray}
 n(\mathbf{x}) &=& |A|^2M^3|\tilde{z}_\mathbf{Q}|^2,\\
\Phi_0(\mathbf{x}) &=& AM^{\frac{3}{2}}\tilde{z}_{\mathbf{Q}},\\
G^{(1)}(\mathbf{x},\mathbf{x'}) &=& A^*A'M^3\tilde{z}_{\mathbf{Q}}^*\tilde{z}_{\mathbf{Q'}},
\end{eqnarray}
with the anomalous average taken to be zero. We reminder the reader that $\tilde{z}_{\mathbf{Q}}$ is defined as the Fourier transform (see above) of $z_{\mathbf{i}}$.
For the results in this section we do not give $G^{(2)}(\mathbf{x},\mathbf{x}^\prime)$ since it is (within the Gaussian approximation) completely determined by $\{\Phi_0(\mathbf{x}),n(\mathbf{x}),G^{(1)}(\mathbf{x},\mathbf{x}^\prime),M(\mathbf{x},\mathbf{x}^\prime)\}$ using (\ref{G2Gaussian}).

\subsection{Inhomogeneous perfect Mott-insulator}\label{InMott}
  Neglecting tunneling, we take the manybody state to be of the product form
 \begin{equation}
 |\Phi_{\rm{M}}\rangle = \prod_{{\mathbf{j}}}\frac{1}{\sqrt{n_{\mathbf{j}}!}}\left(\hat{a}_{{\mathbf{j}}}^\dagger
\right)^{n_{\mathbf{j}}}|0\rangle, \label{nsa}
 \end{equation}
where $n_{{\mathbf{j}}}=\lfloor (\mu-\epsilon_{{\mathbf{j}}})/U\rfloor$ is the integer site occupation, with $\mu$ the
chemical potential  determined by the condition that $\sum_{{\mathbf{j}}}n_{{\mathbf{j}}}=N$.

For this state the condensate fraction and anomalous average are zero
and we obtain  
\begin{eqnarray}
n_{\rm{M}}({\mathbf{x}})&=&|{A}|^2N,\\
G^{(1)}_{\rm{M}}({\mathbf{x}},{\mathbf{x'}})  &=&{A}^*{A'}M^{\frac{3}{2}}\tilde{n}_{{\mathbf{Q}}-{\mathbf{Q'}}}.
\end{eqnarray}
This result shows that  first and second order correlation functions reveal information about the Fourier transform of the density in the system. 

\subsubsection{Application to a super-lattice system}
 \begin{figure}
 \includegraphics[width=3.20in]{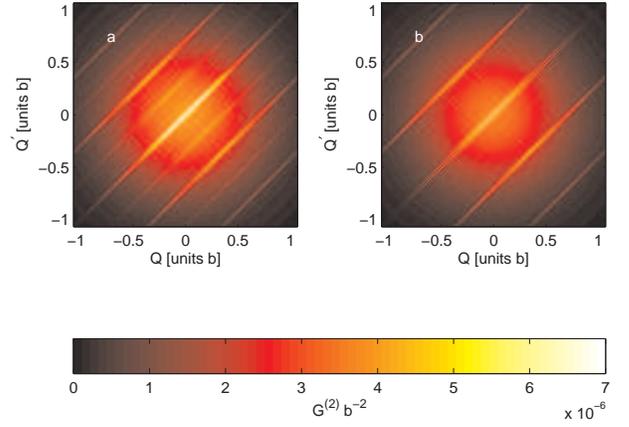}
  \caption{\label{fig:InhomCorr}Comparison of $G^{(2)}$ correlation functions for  1D lattice systems: (a)  Superlattice with periodically modulated density,  (b)
uniformly filled translationally invariant lattice.  Both cases calculated using the inhomogeneous perfect Mott-insulator approach (i.e. tunneling is neglected) with $N=29$ (superlattice), $N=30$ (uniform lattice) and $M=15$.}
  \end{figure}
We now develop a simple application of the inhomogeneous perfect Mott-insulator formalism. We consider a 1D system in which an external potential (i.e. the $\epsilon_j$) causes the many-body state to be of the form given in Eq.~(\ref{nsa}) with 
\begin{equation}
n_j = \left\{    
\begin{array}{cc}
      3, & j\,\rm{even} \\
      1, & j\,\rm{odd} \\
   \end{array}\right. ,
\end{equation}
i.e. a system with a modulated filling. Such an density arrangement can be produced using a superlattice formed by superimposing several optical lattices (e.g. see \cite{Peil2003a,Rey2003b}).
In Fig.~\ref{fig:InhomCorr}(a) we show the second order correlation function for this state [compared to the uniformly filled Mott-insulator state shown in Fig.~\ref{fig:InhomCorr}(b)]. The additional spatial structure due to the density modulation is clearly apparent through the introduction of a new periodicity scale at wavevector $b/2$, i.e. a length scale corresponding to the 2-lattice site modulation in the filling density. Similar conclusions have been reached in a recent experiment analyzing the noise correlations in a  two-color lattice \cite{Guarrerar2008a}.

\subsection{Inhomogeneous ideal gas}
The ground band modes of the (ideal) inhomogeneous Bose Hubbard Hamiltonian satisfy
\begin{equation}
 E_su_{{\mathbf{i}}}^s=-J\sum_{\langle {\mathbf{i}}\, {\mathbf{j}}\rangle}u_{{\mathbf{j}}}^{s}+\epsilon_{{\mathbf{i}}}u_{{\mathbf{i}}}^s,
\end{equation}
where $E_s$ is  the energy eigenvalue  and $u_{\mathbf{i}}^s$ is the amplitude on the ${\mathbf{i}}$-th site  of the $s$-excited mode.
For consistency with the other treatments in this section we take the $s=0$ mode to be that of the condensate (when present) and label its amplitude as $\sqrt{\bar{n}_0}u_{{\mathbf{i}}}^0\to z_{{\mathbf{i}}}$. 
We obtain that
\begin{eqnarray}
\Phi_0({\mathbf{x}}) &=&{A}M^{\frac{3}{2}} \tilde{z}_{\mathbf{Q}},\\
n({\mathbf{x}}) &=& |{A}|^2M^3\left( |\tilde{z}_{{\mathbf{Q}}}|^2+\sum_{s\ne0}{\bar{n}_s}|\tilde{u}^s_{{\mathbf{Q}}}|^2\right),\\
G^{(1)}({\mathbf{x}},{\mathbf{x'}}) &=& {A}^*{A'}M^3\left( \tilde{z}_{{\mathbf{Q}}}^*\tilde{z}_{{\mathbf{Q'}}}+\sum_{s\ne0}{\bar{n}_s}\tilde{u}^{s*}_{{\mathbf{Q}}}\tilde{u}^s_{{\mathbf{Q'}}}\right),
\end{eqnarray}
where $\bar{n}_s$ is the mean thermal occupation of the $s$ mode and the anomalous average is zero.
 
\subsection{Inhomogeneous meanfield decoupling}
The meanfield decoupling approach is immediately extensible to 3D from the form presented in Sec.~\ref{SECDecoupling} by introducing the appropriate generalizations to the parameters introduced earlier, i.e. $n_i\to n_{\mathbf{i}}$, $\alpha_i\to\alpha_{\mathbf{i}}$,  and $\beta_i\to\beta_{\mathbf{i}}$.
In this approach the condensate is identified as
the mean value of the field operator, and we obtain that
\begin{eqnarray}
\Phi_0({\mathbf{x}}) &=& {A}M^{\frac{3}{2}}\tilde{\alpha}_{{\mathbf{Q}}},\\
n({\mathbf{x}}) &=&|{A}|^2 \left\{M^{\frac{3}{2}}\tilde{n}^{I}_{{\mathbf{0}}}+M^3|\tilde{\alpha}_{{\mathbf{Q}}}|^2\right\},\\
G^{(1)}({\mathbf{x}},{\mathbf{x'}})&=& {A}^*{A'} \left\{M^{\frac{3}{2}}\tilde{n}^{I}_{{\mathbf{Q}}-{\mathbf{Q'}}}+M^3\tilde{\alpha}_{{\mathbf{Q}}}^*\tilde{\alpha}_{{\mathbf{Q'}}}\right\},\\
M({\mathbf{x}},{\mathbf{x'}})&=&{A}A'\left\{M^{\frac{3}{2}}\tilde{\beta}_{{\mathbf{Q}}+{\mathbf{Q'}}}-{M^\frac{3}{2}}\widetilde{(\alpha^2)}_{{\mathbf{Q}}+{\mathbf{Q'}}}\right\},
\end{eqnarray}
with $n^{I}_{{\mathbf{j}}}\equiv n_{{\mathbf{j}}}-|\alpha_{{\mathbf{j}}}|^2$ the incoherent density, and $\widetilde{(\alpha^2)}_{{\mathbf{Q}}}$ the Fourier transform of $\alpha^2_{\mathbf{j}}$.

\subsection{Inhomogeneous Bogoliubov approach} 
The extension of Bogoliubov theory to inhomogeneous systems has been given in Refs.~\cite{Rey2003a,Wild2006a}.
Briefly, the discrete Gross-Pitaevskii and Bogoliubov equations
\begin{eqnarray}
\mu z_{\mathbf{j}}&=& -J\sum_{\langle \mathbf{m}
,\mathbf{j}\rangle}z_{\mathbf{m}}+[\epsilon_{\mathbf{j}}+U |z_{\mathbf{j}}|^2]z_{\mathbf{j}},\\
\hbar \omega_s u_{\mathbf{j}}^s&=&-J\sum_{\langle \mathbf{m}
,\mathbf{j}\rangle}u_{\mathbf{m}}^s+[V^{\rm{eff}}_{\mathbf{j}}-\mu]u_{\mathbf{j}}^s-Uz_{\mathbf{j}}^2v_{\mathbf{j}}^s,\\
\hbar \omega_s v_{\mathbf{j}}^s&=&-J\sum_{\langle \mathbf{m}
,\mathbf{j}\rangle}v_{\mathbf{m}}^s+[V^{\rm{eff}}_{\mathbf{j}}-\mu]v_{\mathbf{j}}^s-Uz_{\mathbf{j}}^{*2}u_{\mathbf{j}}^{s},
\end{eqnarray}
need to be solved to describe the system, where $V^{\rm{eff}}_{\mathbf{j}}=\epsilon_{\mathbf{j}}+2U|z_{\mathbf{j}}|^2$ is the effective potential and $s$ labels the  quasi-particle excitations described by the modes $\{u_{\mathbf{j}}^s,v_{\mathbf{j}}^s\}$.

Using these solutions we determine the condensate mode as
\begin{equation}
\Phi_0({\mathbf{x}}) = {A}M^{\frac{3}{2}}\tilde{z}_{{\mathbf{Q}}},
\end{equation}
and the fluctuation operator as
\begin{equation}
\hat{\xi}({\mathbf{x}})={A}M^{\frac{3}{2}}\sum_{s}(\tilde{u}^s_{{\mathbf{Q}}}\hat{\beta}_s-\tilde{v}^{s*}_{{\mathbf{Q}}}\hat{\beta}^\dagger_{s}).
\end{equation}
We also obtain that
\begin{eqnarray}
 n({\mathbf{x}}) &=&|{A}|^2M^3 \left(|\tilde{z}_{{\mathbf{Q}}}|^2+\sum_{s}|\tilde{v}^s_{{\mathbf{Q}}}|^2\right),\\
 G^{(1)}({\mathbf{x}},{\mathbf{x'}})&=&{A}^*{A'}M^3\left(\tilde{z}_{{\mathbf{Q}}}^*\tilde{z}_{{\mathbf{Q'}}}+\sum_{s}\tilde{v}^s_{{\mathbf{Q}}}\tilde{v}_{{\mathbf{Q'}}}^{s}\right),\\
 M({\mathbf{x}},{\mathbf{x'}})&=&-{A}A'M^3\sum_{s}\tilde{u}^s_{{\mathbf{Q}}}\tilde{v}^{s*}_{{\mathbf{Q'}}}.
 \end{eqnarray}

\subsection{Particle-hole corrections to an inhomogeneous Mott Insulator}
 \begin{widetext}
Using perturbation theory  to account for finite tunneling corrections to the inhomogeneous  extension of the perfect Mott-insulator
state (Eq.~(\ref{nsa})), again just as in Sec.~\ref{InMott} the  the condensate fraction and anomalous average are zero
and we obtain  
\begin{equation}
|\Phi_{\rm{M1}}\rangle \approx |\Phi_{\rm{M}}\rangle+\sum_{\langle \mathbf{i},\mathbf{j} \rangle }\frac{J\hat{a}_{\mathbf{i}}^\dagger  \hat{a}_{\mathbf{j}}}{U(1+n_{\mathbf{i}}-n_{\mathbf{j}})+\epsilon_{\mathbf{i}}-\epsilon_{\mathbf{j}}} |\Phi_{\rm{M}}\rangle,
\end{equation} 
for the first order perturbed wavefunction (with $ |\Phi_{\rm{M}}\rangle$ given in (\ref{nsa})) and 
the following expressions for the correlation functions
\begin{eqnarray}
 n_{\rm{M1}}(\mathbf{x}) &=&  n_{\rm{M}}(\mathbf{x})+2 {|A|}^2 \sum_{\langle \mathbf{i},\mathbf{j}\rangle } \cos{(\mathbf{Q}\cdot [\mathbf{R}_{\mathbf{i}}-\mathbf{R}_{\mathbf{j}}])}\frac{J (n_{\mathbf{i}}+1) n_\mathbf{j}}{U(1+n_\mathbf{i}-n_\mathbf{j})+\epsilon_\mathbf{i}-\epsilon_\mathbf{j}}, \label{phnsn}\\
G_{\rm{M1}} ^{(1)} (\mathbf{x},\mathbf{x'})&=& G_{\rm{M}}^{(1)} (\mathbf{x},\mathbf{x'}) +
 A^*A^\prime\sum_{\langle \mathbf{i},\mathbf{j}\rangle } e^{i (\mathbf{Q}-\mathbf{Q}')\cdot\mathbf{R}_{\mathbf{i}}}\left[e^{i\mathbf{Q}\cdot [\mathbf{R}_{\mathbf{i}}-\mathbf{R}_{\mathbf{j}}]}+
 e^{-i\mathbf{Q}^\prime\cdot [\mathbf{R}_{\mathbf{i}}-\mathbf{R}_{\mathbf{j}}]}
 \right] \frac{J(n_{\mathbf{i}}+1) n_\mathbf{j}}{U(1+n_\mathbf{i}-n_\mathbf{j})+\epsilon_\mathbf{i}-\epsilon_\mathbf{j}}. 
  \end{eqnarray} 
\end{widetext}

\section{Conclusions and outlook}
The central purpose of this paper has been to develop a formalism for describing the correlations observed between ultra-cold bosons released from an optical lattice. We  have given validity conditions for a far-field regime, which allowed us to simply relate the correlations measured after expansion to the \emph{in situ} correlations of the system in the lattice. Since the \emph{in situ} state is well-described by the Bose-Hubbard model,  we have developed our formalism for a variety of commonly used methods for solving this model, including: the Bogoliubov method, the meanfield decoupling approach, and the particle-hole perturbative solution about the perfect Mott-insulator state. These methods provide a rather general set of solutions appropriate to a broad range of behavior in the lattice system.
We have also considered a Gaussian version of our formalism that gives general correlation functions of the expanded system in terms of the first order correlation functions. We justify this approach by comparison to our earlier (non-Gaussian) results and show that it provides a complete and accurate description of the results in all regimes considered. Finally, we used the  Gaussian approach to develop a tractable 3D formalism appropriate to inhomogeneous systems -- i.e. relevant to current experiments. The application of this formalism to a variety of experimental regimes will be the subject of future work.

So far experiments have only analyzed the basic correlation properties of bosons in the Mott-insulator regime including second order correlations \cite{Folling2005a,Spielman2007a,Guarrerar2008a} and 
density profiles \cite{Gerbier2005,Spielman2007a,Guarrerar2008a}. Technical noise has so far made  measurements of the superfluid phase difficult, so it would be of interest to see continued refinement of experimental techniques into this regime. Possibly the most interesting region, from the theoretical perspective, is the nature of correlations in the transition region. Preliminary studies of some correlation aspects near the transition have been considered in Ref.~\cite{Spielman2007a}.

Our formalism now presents numerous avenues for future theoretical development. A few such areas of interest are: (i) Characterizing correlation function behavior in the region of the superfluid to Mott-insulator transition. (ii) Investigating the influence of temperature on interacting cases, both in the superfluid and Mott-insulator regimes. (iii) Considering the influence of disorder on the system phase diagram  (e.g. emergence of the Bose-glass phase \cite{Fisher1989a,Buonsante2007a,Lye2007a,Fallani2007a,Guarrerar2008a}) and how signatures of these phases might emerge in correlation functions. 
Some of these problems will require approaches that go beyond those covered here (e.g. may require Monte-Carlo solutions \cite{Kashurnikov2002a,Wessel2004a}), however our results suggest a Gaussian approach will be suitable, i.e. only requiring  knowledge of $\{\Phi_0(\mathbf{x}),G^{(1)}(\mathbf{x},\mathbf{x}'),M(\mathbf{x},\mathbf{x}')\}$.

\section*{Acknowledgments}
Invaluable discussions with P. Buonsante and C.W. Gardiner  are gratefully acknowledged.
 E.T. acknowledges support of  a Top Achiever Doctoral Scholarship.
P.B.B. acknowledges support from the Marsden Fund of New Zealand, the University of Otago and the New Zealand Foundation for Research Science and Technology under the contract NERF-UOOX0703: Quantum Technologies.
 A.M.R. acknowledges support from  the Institute of
Theoretical, Atomic, Molecular and Optical Physics at Harvard
University and Smithsonian Astrophysical observatory.

  \appendix
\section{Projected Density Covariance}\label{secprojcovprops}
Here we examine the properties of the density covariance function constructed from the projected field operator. To be precise we decompose the full field operator into the form
\begin{equation}
\hat{\psi}(\mathbf{x},t)=\hat{\psi}_g(\mathbf{x},t)+\hat{\psi}_e(\mathbf{x},t),\label{psige}
\end{equation}
where $\hat{\psi}_g(\mathbf{x},t)$ is the ground band operator defined in Eq.~(\ref{psiBH}) and $\hat{\psi}_e(\mathbf{x},t)$ describes the modes orthogonal to the ground band. The commutation relations for these operators are
\begin{eqnarray}
 \,[\hat{\psi}_g(\mathbf{x},t),\hat{\psi}_g^{\dagger} (\mathbf{x}^\prime,t)] &=& \Lambda(\mathbf{x},\mathbf{x}^\prime,t),\\
\,[\hat{\psi}_e(\mathbf{x},t),\hat{\psi}_e^{\dagger} (\mathbf{x}^\prime,t)] &=& \delta(\mathbf{x}-\mathbf{x}^\prime)-\Lambda(\mathbf{x},\mathbf{x}^\prime,t),\\
\,[\hat{\psi}_e(\mathbf{x},t),\hat{\psi}_g^{\dagger} (\mathbf{x}^\prime,t)] &=& 0,
\end{eqnarray}
where $\Lambda(\mathbf{x},\mathbf{x}^\prime,t)=\sum_j
w_0( \mathbf{x}-\mathbf{R}_j,t)w_0( \mathbf{x}^\prime-\mathbf{R}_j,t)$.

The experimental observable is the density covariance as defined in Eq.~(\ref{Cov}). Expressed in terms of $\hat{\psi}_g$ and $\hat{\psi}_e$, we obtain
\begin{eqnarray}
C(\mathbf{x},\mathbf{x}^\prime) = C_g(\mathbf{x},\mathbf{x}^\prime) +G^{(1)}(\mathbf{x},\mathbf{x}^\prime)\left\{\delta(\mathbf{x}-\mathbf{x}^\prime)-\Lambda(\mathbf{x},\mathbf{x}^\prime,t)\right\},
\end{eqnarray}
where we have introduced the ground band density covariance
\begin{eqnarray}
C_g(\mathbf{x},\mathbf{x}^\prime) \equiv\langle\hat{n}_g(\mathbf{x})\hat{n}_g(\mathbf{x}^\prime)\rangle-\langle\hat{n}_g(\mathbf{x})\rangle\langle\hat{n}_g(\mathbf{x}^\prime)\rangle,
\end{eqnarray}
with $\hat{n}_g(\mathbf{x})\equiv \hat{\psi}^\dagger_g(\mathbf{x})\hat{\psi}_g(\mathbf{x})$, and have assumed the excited bands to be in their vacuum state so that $\langle\hat{\psi}^\dagger_e(\mathbf{x}^\prime)\hat{\psi}_e(\mathbf{x}^\prime)\rangle=0$.

\section{Finite resolution}
The issue of finite imaging resolution and the effect of taking a column density has been dealt with in reference \cite{Folling2005a}, however we review the basic considerations here for completeness.

For the case of absorption imaging, the measured density is given by 
\begin{equation}
\hat{n}_{\rm{m}}(x,y)\approx \int dx'dy'\,{\mbox{PSF}}(x',y')\int dz\,\hat{n}(x-x',y-y',z),\label{n_meas}
\end{equation}
where the integral along $z$ accounts for taking column density  and the point spread function, ${\mbox{PSF}}(x,y)$, accounts for the limited optical and detector resolution. The point spread function can be approximated as a normalized Gaussian function of widths $\sigma_x$ and $\sigma_y$ along the $x$ and $y$ directions respectively, i.e. ${\mbox{PSF}}(x,y)\approx\prod_{j=x,y}\left\{\exp(-x_j^2/2\sigma_j^2)/\sqrt{2\pi}\sigma_j\right\}$.
We note that for the case where the atoms are counted directly, e.g. using a micro-channel detector plate in a meta-stable helium system, we can associate a 3D point spread function for the measurements which includes the (2D) spatial resolution of the detector and the time resolution.

The general situation of interest is when a peaked feature occurs in a correlation function. To understand the effects of imperfect resolution we consider a 1D example to develop the basic ideas.
Take the peaked function to be of the form 
\begin{equation}
f_{\rm{peak}}(x)=h\exp[-x^2/2(\Delta x)^2],
\end{equation}
 with feature height $h$. After convolution with the (1D) point spread function $\mbox{PSF}(x)=\exp(-x^2/2\sigma_x^2)/\sqrt{2\pi}\sigma_x$ we obtain the \emph{measured} quantity
\begin{equation}
f_{\rm{peak}}^{\rm{meas}}(x)=h\Lambda_x\exp\left(-\frac{x^2}{2W_x^2}\right),
\end{equation}
where
\begin{eqnarray}
\Lambda_x &=& \frac{1}{\sigma_x\sqrt{\frac{1}{(\Delta x)^2}+\frac{1}{(\sigma_x)^2}}},\\
W_x &=& \sqrt{(\Delta x)^2+\sigma_x^2},
\end{eqnarray}
are the resulting height ratio and width parameters respectively.

Several approximate limits for the effects of imaging on the height and width of the features in $f_{\rm{peak}}^{\rm{meas}}(x)$ are easily found:
\subsubsection{perfect resolution: $\Delta x\gg \sigma_x$}
In this limit the resolution is much better than the characteristic size of the peak feature and we have
\begin{eqnarray}
W_x &\approx & \Delta x,\\
\Lambda_x & \approx& 1.
\end{eqnarray}
In this case the features of the peak are accurately revealed by the imaging.

\subsubsection{limited resolution: $\Delta x\ll \sigma_x$}
In this limit the resolution is larger than the characteristic size of the peak feature and we have
\begin{eqnarray}
W_x &\approx & \sigma_x,\\
\Lambda_x & \approx& \frac{\Delta x}{\sigma_x}\ll 1. \label{Wj_badresol}
\end{eqnarray}
Thus the measured peak feature is strongly altered by imaging, in particular the width is that of the imaging resolution and the feature height is reduced according to the ratio of the feature size to imaging resolution.

\subsubsection{no resolution}
For the lattice system there is usually a limit to the ratio $\Delta x/\sigma_x$ appearing in Eq.~(\ref{Wj_badresol}). The finite number of lattice sites makes the smallest feature  size $(\Delta x)_{\min}=l_b/M$ where $l_b = \hbar bt/m$ is the periodic length scale of the expanded system (also see Eq.~(\ref{l_periodic})). In practice the largest value of the imaging resolution  is $(\sigma_j)_{\max}=l_b$, since beyond this length scale the underlying periodicity in the system will cause the peak to reoccur.
Thus, in the  \emph{no resolution} limit we have that (from Eq.~(\ref{Wj_badresol}))
\begin{equation}
\Lambda_x\approx \frac{1}{M}.
\end{equation}

\subsection{Generic experimental situation}
Generally experiments (e.g. Refs \cite{Folling2005a,Rom2006a,Spielman2007a}) have operated in the regime where $\Delta x_j\approx l_b/M\ll\sigma_j\ll l_b$ for $j=x,y$ (in the imaging plane) and taking a column density (i.e.~\emph{no resolution} limit)  along $z$. Most importantly this means the height of peaks in the correlation function are reduced by a factor of
\begin{equation}
\prod_j\Lambda_j = \frac{l_b^2}{M^3\sigma_x\sigma_y}.
\end{equation}
 A more complete treatment yields an additional factor of $1/4\pi$ \footnote{Since the correlation functions are produced by convolving two density images $\{\sigma_x,\sigma_y\}$ will be twice those values characterizing the resolution of the imaging system.}. The widths of the peaks will be given by $\sigma_x$ and $\sigma_y$ in the $x$ and $y$ directions respectively.

\bibliography{correl}
\bibliographystyle{apsrev}

\end{document}